\documentclass{article}
\usepackage{graphicx}
\usepackage{latexsym}
\usepackage{amsmath,amssymb,amsfonts}
\usepackage{bm}
\usepackage{color}
\usepackage[pdftex]{geometry}
\newcommand{\ket}[1]{\left| #1 \right\rangle}
\newcommand{\bra}[1]{\left\langle #1 \right|}

\begin{document}

\title{NMR Contributions to the study of Quantum Correlations}


\author{Isabela A. Silva$^{1}$, Jefferson G. Filgueiras$^{1}$, Ruben Auccaise$^{2}$, Alexandre M. Souza$^{3}$,
\\Raimund Marx$^{4}$, Steffen J. Glaser$^{4}$, Tito J. Bonagamba$^{1}$,
\\Roberto S. Sarthour$^{3}$, Ivan S. Oliveira$^{3}$, Eduardo R. deAzevedo$^{1}$}

\maketitle

\noindent{\small{\it $^1$ Instituto de F\'isica de S\~{a}o Carlos, Universidade de S\~{a}o Paulo,
P.O.~Box 369,  S\~{a}o Carlos, 13560-970 S\~{a}o Paulo, Brazil \newline
$^2$ Universidade Estadual de Ponta Grossa, Dept Fis, BR-84030900 Ponta Grossa, Parana, Brazil \newline
$^3$ Centro Brasileiro de Pesquisas F\'isicas, Rua Dr. Xavier Sigaud 150,
Rio de Janeiro, 22290-180  Rio de Janeiro, Brazil \newline
$^4$ Department of Chemistry, Technische Universit\"{a}t M\"{u}nchen, Lichtenbergstr. 4, 85747, Garching, Germany \newline
}}

\date{}

\begin{abstract}
\noindent In this chapter we review the contributions of Nuclear Magnetic Resonance to the study of quantum correlations, including its capabilities to prepare initial states, generate unitary transformations, and characterize the final state. These are the three main demands to implement quantum information processing in a physical system, which NMR offers, nearly to perfection,  though for a small number of qubits. Our main discussion will concern liquid samples at room temperature. 
\end{abstract}
\baselineskip=4ex

\section{NMR fundamentals}

\subsection{Classical NMR}
Nuclear Magnetic Resonance (NMR) was first reported in 1939 by Isidor Rabi and co-workers, as a method to measure nuclear magnetic moments, a work inspired by the earlier Stern-Gerlach experiment \cite{rabi}. For that work Rabi received the Nobel Prize in Physics in 1944. In 1946 Felix Bloch and Edward Purcell \cite{bloch,purcell} demonstrated, independently, the NMR phenomenon in matter, solid and liquid, and reached an adequate parametric mathematical formulation known as the {\em Bloch Equations}. For that work they received the Nobel Prize in Physics in 1952.  In 1950  a giant step was given by Erwin Hahn, a step that would turn NMR, in the years to come, into one of the main experimental techniques in Physics, Chemistry and Biology with a revolutionary application to Medicine. Hahn discovered the phenomenon of {\em spin echoes} \cite{hahn}, inaugurating {\em Pulsed NMR}. For further developments and applications of pulsed NMR, the Nobel Prize in Chemistry in 1991 was granted to Richard Ernst \cite{ernst}. In 2002 another NMR Nobel Prize in Chemistry was awarded to Kurt Wuthrich \cite{wuthrich} and, in 2003, the Nobel Prize in Medicine went to Paul Laterbur and Peter Mansfield for the discovery of the NMR imaging technique \cite{laterbur,mansfield}. Therefore, since its discovery, NMR has been awarded five Nobel Prizes, three of them due to contributions after the discovery of its pulsed version. 

In a basic NMR experiment \cite{slichter}, an ensemble of nuclear magnetic moments are subject to a magnetic field given by:

\begin{equation}\label{eq1}
{\bf B}(t) = B_0{\bf k} + B_1\left\{\cos (\omega t){\bf i} + \sin (\omega t){\bf j}\right\}
\end{equation}In this expression, $B_0$ is the magnitude of a homogeneous static magnetic field, typically of the order of 10 Tesla, whereas $B_1$ is the magnitude of a {\em radiofrequency field} (RF), typically four to five orders of magnitude below $B_0$. Therefore, $B_1$ can be considered a perturbation over $B_0$. It is worth mentioning that, although $B_0$ is considered homogeneous in the above equation, the description of the NMR phenomenon necessarily include a field {\em inhomogeneity} \cite{slichter}.

In the classical description of NMR the field described by Eq.(\ref{eq1}) interacts with the nuclear magnetization producing a torque on it. The magnetization rotates with a characteristic frequency given by $\omega_0/2\pi = (\gamma_n/2\pi)B_0$, where $\gamma_n$ is a nuclear parameter called {\em gyromagnetic ratio}, different for each isotopic specimen. This frequency, called the {\em Larmor frequency} of the system, can range from a few to many hundreds of MHz. The presence of time-dependent terms in the field complicates the description of the time evolution of the  magnetization in the laboratory frame. Fortunately, due to the special geometrical arrangement, it is possible to make a transformation to a rotating frame in which the total field is static \cite{slichter}. By adding the relaxation terms, we arrive at the Bloch Equations in the rotating frame, which can be conveniently written in the matrix form:

\begin{equation}\label{eq2}
\frac{\partial {\bf M}}{\partial t}+\tilde{{\bf A}}{\bf M} = {\bf f}
\end{equation}where:
\begin{equation}
\tilde{{\bf A}} = \left(
    \begin{array}{ccc}
    1/T_2      &    - \Delta\omega    &     0     \\
    +\Delta\omega  &       1/T_2       &   - \omega_1     \\
      0  &      \omega_1    &      1/T_1
    \end{array}
    \right); \;\;\;
    {\bf M} = \left(
    \begin{array}{c}
    M_x      \\
    M_y      \\
    M_z
    \end{array}
    \right);\;\;\;
     {\bf f} = \left(
    \begin{array}{c}
    0      \\
    0      \\
    M_0/T_1
    \end{array}
    \right)
\end{equation} In this equation, $M_0$ is the equilibrium magnetization, $T_1$ and $T_2$ are, respectively, the longitudinal and transverse relaxation times, and $\Delta\omega = \omega -\omega_0$ is the {\em offset} between the RF and the resonance frequencies. The resonance condition is given by $\omega = \omega_0$. Finally, $\omega_1 =\gamma_nB_1$ is the rotating frequency of the magnetization about $B_1$ in the rotating frame. From the microscopic point of view, the longitudinal relaxation describes processes in which the nuclear spins interact with a bath and decay towards equilibrium by releasing energy (heat). The transverse  relaxation is more subtle: it describes processes in which the nuclear magnetization loses coherence (quantum and classical) due to the random interactions between the nuclear magnetic moments; the energy is conserved and there is no heat transfer to the bath in the process.

The general solution of Eq.(\ref{eq2}) is:
\begin{equation}\label{eq3}
{\bf M}(t)={\bf M}_\infty + \exp (-\tilde{{\bf A}}t)\times \left\{{\bf M}(0)-{\bf M}_\infty \right\}
\end{equation} where $ {\bf M}_\infty $  is the stationary solution:

\begin{equation}\label{eq4}
{\bf M}_\infty = \tilde{{\bf A}}^{-1}{\bf f}
\end{equation}We see that the dynamics of the magnetization is governed by the exponential operator on the transient (second) term of Eq.(\ref{eq3}). In spite of its apparent simplicity, it is not an easy task to obtain an analytical expression for the magnetization for arbitrary $T_1$, $T_2$ and $\Delta\omega$ \cite{Bain}. Figure \ref{figmags} shows the trajectory of the magnetization, calculated numerically from Eq. (\ref{eq3}), for three different regimes of relaxation: slow, intermediate and fast, and $\Delta\omega\neq 0$. Figure \ref{figtime} shows the time evolution of the components of the magnetization with transient and stationary regimes. We see that at resonance, $M_x = 0$ and the magnetization rotates only about the field $B_1$. We also see that in the regime of fast relaxation and/or low RF power (small $B_1^2$), $M_z\approx M_0$, which means the system absorbs RF energy and releases it very fast as heat to the bath. On the other hand, in the regime of negligibly slow relaxation, the solutions of Bloch equations can be found easily:

\begin{figure}
\begin{center}
\includegraphics[scale=0.3]{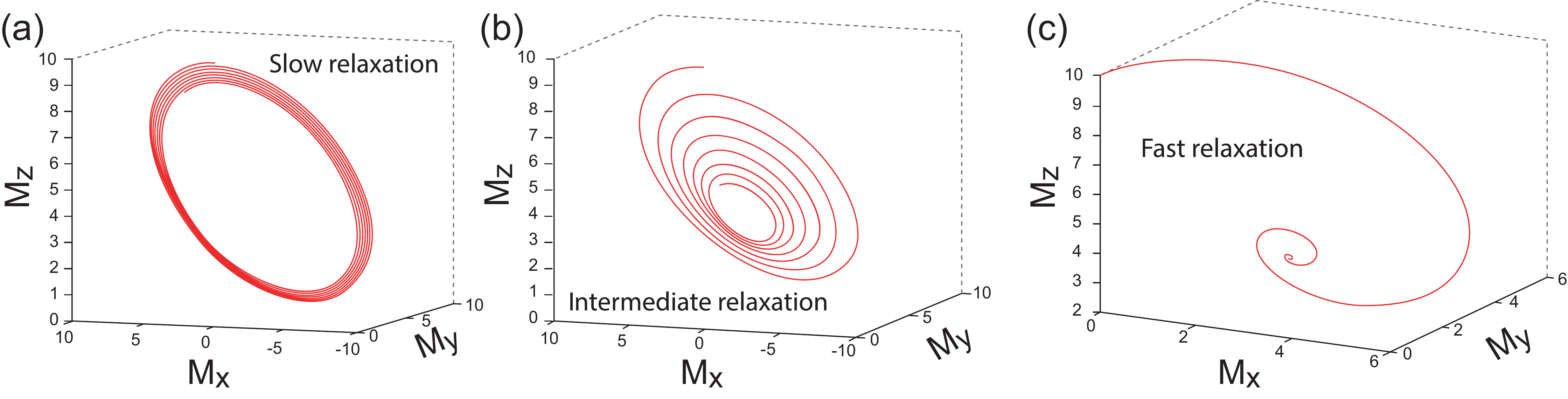}
\caption{Trajectories of the magnetization for different relaxation regimes.}
\label{figmags}
\end{center}
\end{figure}

\[
M_x(t)=-2M_0\frac{\omega_1\Delta\omega}{\Omega^2}\sin^2\left(\frac{\Omega t}{2}\right)
\]
\begin{equation}\label{eq5}
M_y(t)=-M_0\frac{\omega_1}{\Omega}\sin\left(\frac{\Omega t}{2}\right)
\end{equation}
\[M_z(t)=M_0\left[1-2\frac{\omega_1^2}{\Omega^2}\sin^2\left(\frac{\Omega t}{2}\right)\right] \]where $\Omega=\sqrt{\Delta\omega^2+\omega_1^2}$. As long as $T_1$ is much longer than the duration of a RF pulse, we can consider the system isolated. In this regime, using the above results we can calculate the work done by a RF pulse of duration $\tau$ to rotate the magnetization. This will be simply the difference between final and initial internal energy:

\begin{equation}\label{eq6}
W=U_{final}-U_{initial}=-B_0M_z(\tau)+B_0M_0=2B_0M_0\frac{\omega_1^2}{\Omega^2}\sin^2\left(\frac{\Omega \tau}{2}\right)
\end{equation}

\begin{figure}
\begin{center}
\includegraphics[scale=0.3]{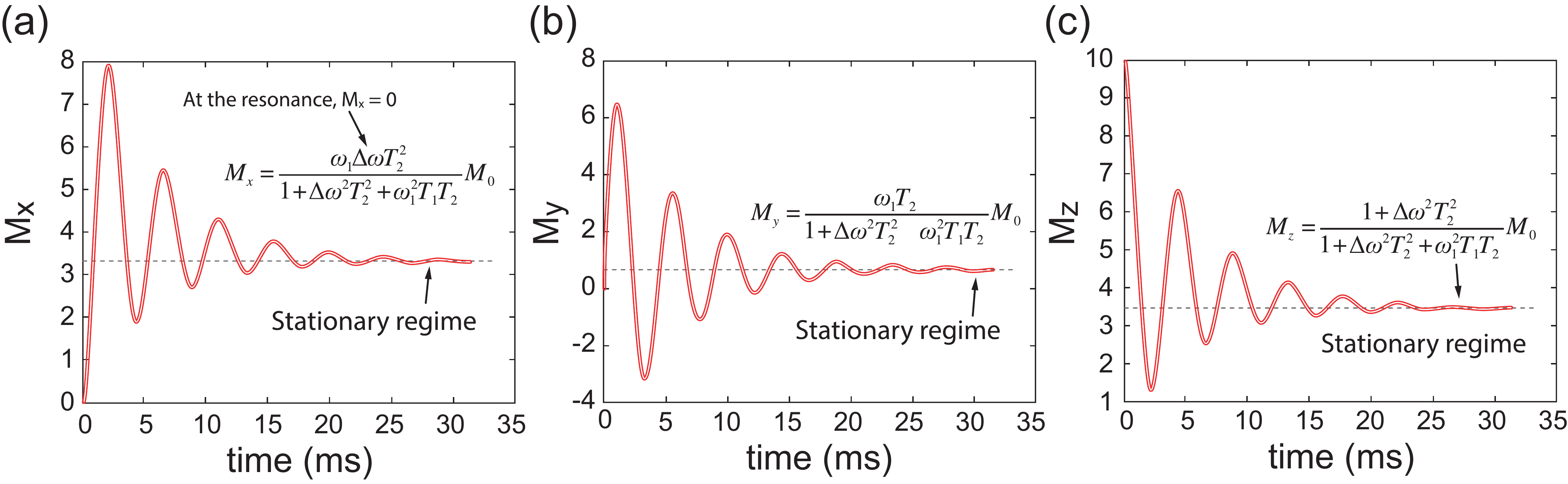}
\caption{Time evolution of the magnetization with transient and stationary regimes.}
\label{figtime}
\end{center}
\end{figure}

\subsection{Quantum NMR}
NMR has also a very cool quantum description. For an isolated spin-1/2, which encodes a quantum bit (qubit) in quantum information processing experiments, under a rotating frame hamiltonian \cite{slichter}:

\begin{equation}\label{eq7}
H= -\frac{1}{2}\hbar\Omega\sigma_u
\end{equation}where $\sigma_u$ is the component of the spin along the direction of the effective field:

\begin{equation}\label{eq8}
\sigma_u = \frac{\Delta\omega}{\Omega}\sigma_z-\frac{\omega_1}{\Omega}\sigma_x
\end{equation}Assuming that a spin is initially in the state $|\uparrow\rangle$, aligned with $B_0$, a RF pulse of duration $\tau$ takes the state to:

\begin{equation}\label{eq9}
|\psi(\tau)\rangle =e^{-i(\Omega\tau/2)\sigma_u}|\uparrow\rangle =\left\{I\cos\left(\frac{\Omega\tau}{2}\right)-
i\frac{\Delta\omega}{\Omega}\sin\left(\frac{\Omega\tau}{2}\right)\right\}|\uparrow\rangle +
\frac{\omega_1}{\Omega}\sin\left(\frac{\Omega\tau}{2}\right)|\downarrow\rangle
\end{equation}The expectation value of the $z$ component, and energy at $\tau$ are:

\begin{equation}\label{eq9a}
\langle\sigma_z\rangle(\tau )=\langle\psi (\tau)|\sigma_z|\psi (\tau )\rangle  =1-2\frac{\omega_1^2}{\Omega^2}\sin^2\left(\frac{\Omega\tau}{2}\right)\end{equation} and
\begin{equation}\label{eq10}
E_f=-\frac{\hbar\omega_0}{2}+\hbar\omega_0\frac{\omega_1^2}{\Omega^2}\sin^2\left(\frac{\Omega\tau}{2}\right)
\end{equation}Therefore, the work produced by the pulse is:

\begin{equation}\label{eq11}
W=E_f-E_i=\hbar\omega_0\frac{\omega_1^2}{\Omega^2}\sin^2\left(\frac{\Omega\tau}{2}\right)
\end{equation}which is the same as the classical result, Eq.(\ref{eq6}), if we remember that for a spin-1/2, $\hbar\omega_0=2\mu_NB_0$, where $\mu_N$ is the nuclear magneton.

The above calculation was made for an isolated spin. If we allow a thermal contact with a bath at equilibrium the initial magnetization will be $\overline{\langle\sigma_z\rangle}_0$. Besides, if there are fluctuations in the work performed by the pulse, the {\em average work} will be given by \cite{simeao}:

\begin{equation}\label{eq12}
\langle W\rangle=2\overline{\langle\sigma_z\rangle}_0B_0\frac{\omega_1^2}{\Omega^2}\sin^2\left(\frac{\Omega\tau}{2}\right)
\end{equation}That is, in the presence of fluctuations, the quantum average work equals the classical expression in the absence of relaxation. An equivalent quantum expression for the average work in the presence of relaxation, is still lacking in the literature.

NMR samples for quantum information processing applications are mainly liquids, typically 0.6 cc, in a glass tube at room temperature. The tube is positioned at the center of a RF coil, which is in turn placed in a static magnetic field. After a few seconds the nuclear spins reach thermal equilibrium in the field, and the initial state of the system is given by the thermal density matrix:

\begin{equation}\label{eq13}
\rho_{eq}=\frac{e^{-H/k_BT}}{{\cal Z}}
\end{equation}where ${\cal Z}$ is the partition function. The energy scale of the hamiltonian is $\mu_NB_0 \approx 10^{-6}$ eV, much smaller than thermal energy, $k_BT\approx 0.01$ eV for room temperature. Therefore, the equilibrium density matrix can be expanded to first order:

\begin{equation}\label{eq14}
\rho_{eq}\approx \frac{I}{2^N}-\frac{H}{2^Nk_BT}
\end{equation}where $I$ is the identity matrix, ${\cal Z}\approx 2^N$ and $N$ is the number of quantum bits (qubits) in the system. The application of RF to the sample can generally be represented by an unitary transformation, $U$, of the density matrix:

\begin{equation}\label{eq15}
\rho^\prime = U\rho_{eq}U^\dag = \frac{I}{2^N}-\frac{UHU^\dag}{2^Nk_BT}
\end{equation}Therefore, NMR state transformations turn out to be hamiltonian transformations. If after the transformation the mean value of the spin component $\sigma_u$ is measured, we get the nuclear magnetization in the direction $u$:

\begin{equation}\label{eq16}
M_u={\rm Tr}\left\{\sigma_u\rho^\prime\right\}=-\frac{1}{2^Nk_BT}{\rm Tr}\left\{\sigma_uUHU^\dag\right\}
\end{equation} This result is extremely important for NMR quantum information processing: the measured signal is not sensitive to the first term, proportional to the identity. This term represents a huge amount of noise, but which is simply invisible to NMR! By combining unitary transformations and space or time averages, the state (\ref{eq13}) can be transformed to \cite{oliveira}:

\begin{equation}\label{eq17}
\rho_{pp} = \frac{1-\epsilon}{2^N}I+\epsilon |\psi\rangle\langle\psi|
\end{equation}where $\epsilon \approx \hbar\omega_0/k_BT\approx 10^{-6}$ for liquid samples at room temperature at magnetic field of the order of 10 Tesla, and $|\psi\rangle$ a pure quantum state. This is a non-equilibrium state known as {\em pseudopure state}. It is so-called because under unital operations it transforms as a pure state. Besides, if a measurement of $\langle\sigma_u\rangle$ is performed for this state, the detected signal will come only from the $\epsilon |\psi\rangle\langle\psi|$ part of the pseudopure state:

\begin{equation}\label{eq18}
M_u=\epsilon{\rm Tr}\left\{\sigma_u|\psi\rangle\langle\psi|\right\}
\end{equation}The fact that $\epsilon$ is small is immaterial for both, transformation of states and detection of NMR signal. In a standard NMR quantum information processing task, the initial density matrix is the pseudopure state:

\begin{equation}\label{eq19}
 \rho_{pp}^0 = \frac{1-\epsilon}{2^N}I+\epsilon |00\cdots 000\rangle\langle 00\cdots 00|,
\end{equation}a protocol represented by unitary transformations $U_1,U_2,\cdots U_N$ is applied:

\begin{equation}\label{eq20}
 \rho_{pp}^\prime = \frac{1-\epsilon}{2^N}I+\epsilon \left\{U_N\cdots U_2U_1|00\cdots 0\rangle\langle 00\cdots 00|U_1^\dag U_2^\dag\cdots U_N^\dag\right\},
\end{equation}and the NMR signal is measured:

\begin{equation}\label{eq21}
M_u = \epsilon {\rm Tr}\left[\sigma_u\left\{U_N\cdots U_2U_1|00\cdots 0\rangle\langle 00\cdots 00|U_1^\dag U_2^\dag\cdots U_N^\dag\right\}\right]
\end{equation}By performing multiple measurements, the full density matrix can be reconstructed, the process called {\em Quantum State Tomography} \cite{oliveira}. Among the many examples in the literature on the application of this general procedure are NMR testing of Bell-inequality violation \cite{alexandre1}, violation of Leggert-Garg inequality \cite{alexandre2}, and an NMR quantum complemetarity principle study \cite{ruben1, Marx1}.

\section{NMR Classical and Quantum Correlations}\label{NMRentang}
The NMR observable is the transverse nuclear magnetization, following a sequence of RF pulses. It is a classical quantity. Suppose a $90^o$ pulse is applied along the $x$ axis to an ensemble of nuclear magnetic moments initially at thermal equilibrium with a static and homogeneous magnetic field along the $z$ axis. If the initial magnetization is $M_0$, after the pulse the magnetization will be $M_y =M_0$ and $M_z=0$. Neglecting longitudinal relaxation, in the absence of field inhomogeneity and spin-spin interactions, the magnetization would rotate permanently about the $z$ direction, maintaining its initial magnitude $M_0$. That is, all the magnetic moments in the sample, something like $10^{18}$ along the sensitive region of the sample holder, would rotate in perfect synchronism keeping their relative initial phase difference. This is a classical NMR coherent state. However, because field inhomogeneities and spin-spin interactions are always present, the initial coherent state $M_y=M_0$ will dephase, first due to field inhomogeneity and then due to spin-spin interaction. The first effect is reversible in a spin-echo experiment, but the second is not. It is worth mentioning that until quite recently the debate whether the spin-echo phenomenon violates the Second Law of Thermodynamics could be found in the literature \cite{waugh,anastopoulus}.

Those two independent relaxation phenomena are well described applying the operator-sum formalism \cite{NielsenChuang}, which is possible since every transformation that is given by a complete positive map admits a representation like
\begin{eqnarray}
\sigma(t) = \sum_kE_k(t)\sigma(0)E_k^\dagger(t),
\end{eqnarray}
with $E_k(t)$ being the Kraus operators satisfying
\begin{eqnarray}
\sum_kE_k^\dagger(t)E_k(t)=1.
\end{eqnarray}

The transverse relaxation process in liquid state NMR is exactly analogous to the quantum 
information's phase damping (PD) channel, mathematically described as
\begin{eqnarray}
E_1=\sqrt{1-\frac{q(t)}{2}}\mathbb{I},\quad E_2=\sqrt{\frac{q(t)}{2}}\left[
\begin{array}{cc}
1 & 0 \\
0 & -1 \end{array}\right],
\end{eqnarray}
where $q(t) = 1 - e^{-t/T_2}$, and $T_2$ being the NMR transversal relaxation characteristic time.

Moreover, the longitudinal relaxation is also known as Generalized Amplitude Damping (GAD) channel and has the mathematical description
\begin{eqnarray}
E_1 = \sqrt{p} \left[\begin{array}{cc}
1 & 0 \\
0 & \sqrt{1-\gamma} \end{array} \right],\, E_2 = \sqrt{p} \left[\begin{array}{cc}
0 & \sqrt{\gamma} \\
0 & 0 \end{array} \right] \\
E_3 = \sqrt{1-p}\left[\begin{array}{cc}
\sqrt{1-\gamma} & 0 \\
0 & 1 \end{array} \right], \, E_4 = \sqrt{1-p}\left[\begin{array}{cc}
0 & 0 \\
\sqrt{\gamma} & 0 \end{array} \right], \nonumber
\end{eqnarray}
where $\gamma = 1 - e^{-t/T_1}$, $p \approx (1-\alpha)/2$, $\alpha = \hbar\omega_L/k_BT$ and $T_1$ the NMR longitudinal relaxation characteristic time.

\subsection{NMR Entanglement}
NMR contributions to quantum information processing appeared right after the discovery of pseudopure states in 1997 \cite{gershenfeld,cory}. In the year after, the question of NMR entanglement was raised \cite{browstein}, and further elaborated \cite{linden}. From Eq. (\ref{eq18}) we see that the signal measured from a peudopure state is proportional to spins in a single quantum state. Therefore if $|\psi\rangle$ is entangled, the NMR signal will bring the signature of spins in an entangled state, as demonstrated by quantum state tomography in various works \cite{fatima}. However, if we consider the whole density matrix, Eq.(\ref{eq17}), for $\epsilon$ below a threshold, the state will be separable. As an example consider two qubits in a pseudo-entangled state. From (\ref{eq18}):

\begin{equation}\label{eq22}
\rho_{pp} = \left(
    \begin{array}{cccc}
    (1+\epsilon)/4      &           0          &     0               &  0     \\
      0                 &      (1+\epsilon)/4  &    -\epsilon/2      &  0    \\
      0                 &       -\epsilon/2    &  (1+\epsilon)/4     &  0     \\
      0                 &           0          &     0               & (1+\epsilon)/4
    \end{array}
    \right)
    \end{equation}
 Calculating the eigenvalues of the partially transposed matrix, we obtain $\lambda_1=\lambda_2=\lambda_3=(1+\epsilon)/4$ and $\lambda_4=(1-3\epsilon)/4$. Therefore, according to Peres criterium \cite{peres}, the state will be entangled for $\epsilon >1/3$, much above that for room temperature liquid-state samples.

The question of NMR entanglement is quite subtle, and the whole thing has to do with the value of $\epsilon$, which does not affect neither the unitary transformations, nor the measured signal besides its intensity. For instance, in Ref. \cite{alexandre1} the NMR protocol for an experiment of Bell inequality violation produces a result which is indistinguishable from those obtained in a quantum optics experiment or quantum mechanical prediction. For that comparison the NMR data are normalized by the factor $\epsilon$. Of course, if the normalization is not performed the curves cannot be compared with each other. The equivalent procedure in a quantum optics experiment is called {\em post-selection}, in which pairs of entangled photons are selected out from the total number of detected particles. If in a such experiment one pair of entangled photons is detected with probability $10^{-6}$ we would have the analog of a NMR experiment \cite{Fahmy}.

It has been shown that some of the aspects of entangled states cannot be tested by NMR, such as nonlocality 
\cite{MenicucciC}. However, the fact that NMR is capable of implementing all the basic steps for quantum information processing, state preparation, unitary evolution and quantum state tomography, makes the technique a unique laboratory system to test quantum information protocols in small systems. A non-exhaustive compilation of NMR earlier works involving entanglement can be found in \cite{oliveira}. For a recent compilation of NMR Quantum Information Processing from various research groups in the World, see \cite{rs}.

One key aspect of NMR to study quantum correlations, in general, and entanglement, in particular, is the ability to prepare very specific initial states. In a two-qubit state scenario, an important class of states are the Bell diagonal states, mathematically written as
\begin{eqnarray}
\rho = \frac{1}{4}\left[\mathbb{I} + \sum_{i = 1}^{3} c_i(\sigma_i\otimes\sigma_i)\right],
\label{m32}
\end{eqnarray}
where a full description only depends on the correlation triple $\vec{c} = \{c_1, c_2, c_3\}$, with $c_i = \textrm{Tr}\{\rho(\sigma_i\otimes\sigma_i)\}$. In Ref. \cite{chuang}, this class of states was first prepared in NMR, and the pulse sequence implemented, described in Fig. \ref{DSC12}-A,  can be easily changed to produce a desired Bell diagonal state. 

\begin{figure}
\begin{center}
\includegraphics[scale=0.35]{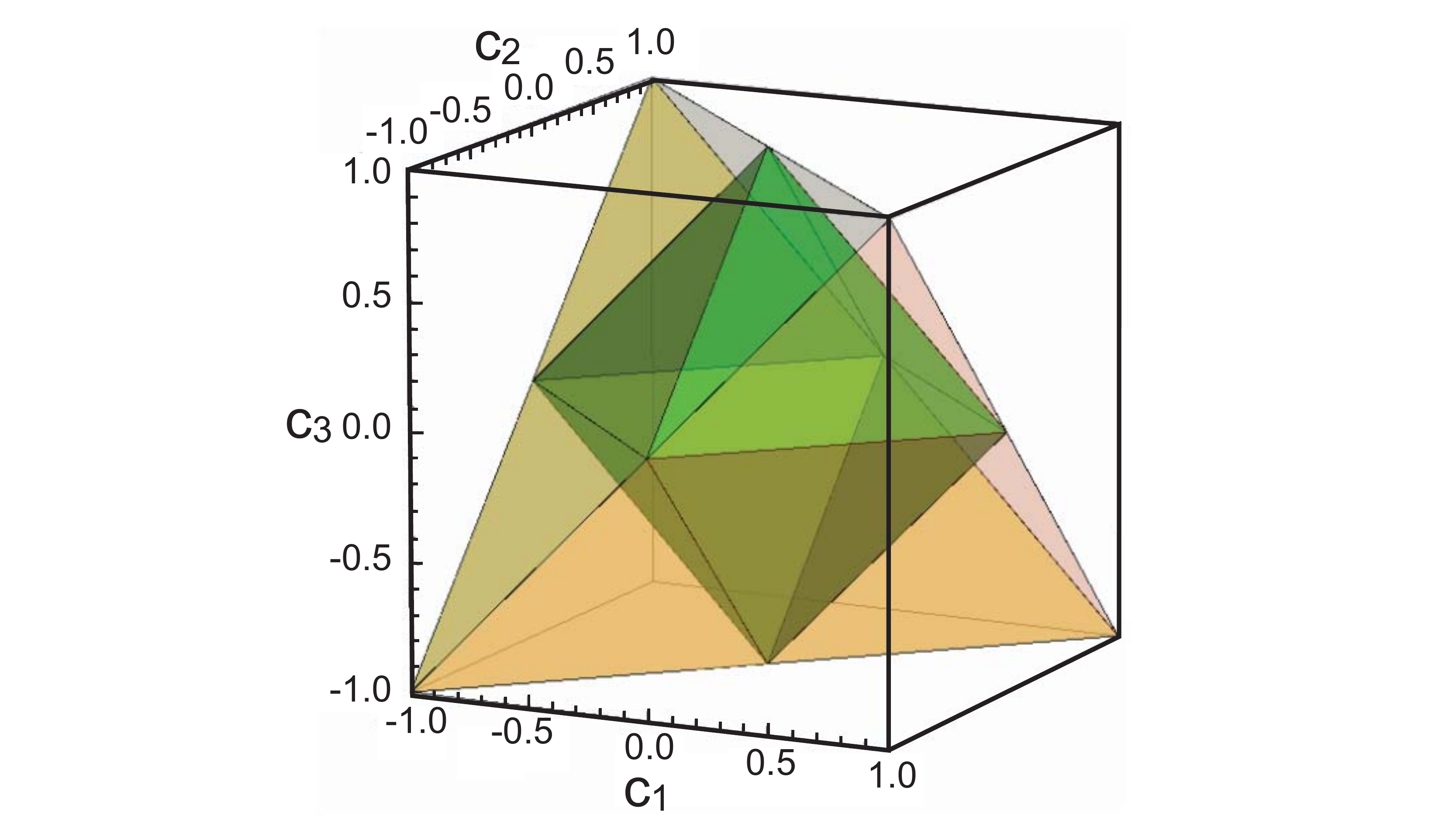}
\caption{Bell diagonal states geometry. The outside tetrahedron is the set of physical Bell diagonal states. The inside octahedron, characterized by $|c_1|+|c_2|+|c_3|\leq 1 (\lambda_l\leq1/2)$ are the separable Bell diagonal states. The physical states outside the octahedron are entangled states. The maximally entangled Bell states are in the tetrahedron vertices. The classical states (diagonal in the product basis) belong to the cartesian axes.}
\label{tetrahedron}
\end{center}
\end{figure}

Fig. \ref{tetrahedron} shows all physical Bell states. The states in the vertices are maximally entangled states. States inside the octahedron are separable, however presenting non-zero quantum correlations, measured by quantum discord, the subject of the next section.

In a more general scenario, this class of two-qubit states can be generalized to
\begin{eqnarray}
\rho = \frac{1}{2^N}\left[\mathbb{I}^{\otimes N} + \sum_{i = 1}^{3} c_i(\sigma_i^{\otimes N})\right],
\label{m3N}
\end{eqnarray}
which was named $M^3_N$ states \cite{bromley}, meaning N-qubit states with maximally mixed marginals. This general case will present similar properties as the Bell diagonal ($M^3_2$) one, as we shall discuss in the next sections.  

\subsection{NMR Discord}
{\em Correlation} is a key concept in statistics. Two random variables are said to be correlated when knowledge about one of them can be gained by measuring the other. Since the seminal work of Shannon \cite{shannon}, information is quantified by {\em entropy}, a quantity which appears with different names in diverse contexts: thermodynamical entropy, statistical entropy, Shannon entropy, von Neumann entropy. The concept was proposed by Rudolf Clausius (1822-1888) as a ``measure of the energy'' in a thermodynamical system not available for the realization of work. The statistical entropy connects microscopic dynamics with macroscopic thermodynamical quantities. Shannon entropy is measured in bits: the entropy of one bit of information is equal to 1. But it is only the von Neumann entropy which captures the correlations present in quantum states.

It is possible to quantify quantum correlation, even in thermal systems in the presence of noise. Let us recall the definition of Shannon entropy, associated to two dichotomic random variables $X$ and $Y$ which can take the values $\{x,y\}$ with probabilities $\{p_x,p_y\}$:

\begin{equation}\label{eq24}
S_X=-\sum_ip_x\log_2p_x\;\;{\rm and}\;\; S_Y=-\sum_ip_y\log_2p_y
\end{equation}$S_{X(Y)}$ quantifies our uncertainty about $X(Y)$: the larger the entropy, the less we know about the system. If $X$ and $Y$ are correlated, a measurement of $Y$ will yield information about $X$. Then, our knowledge about $X$ must be ``updated''; the new entropy of $X$ after we got to know $Y$ is represented by $S_{X|Y}$:

\begin{equation}\label{eq25}
S_{X|Y}=S_{X,Y}-S_Y
\end{equation}where:

\begin{equation}\label{eq27}
S_{X,Y}=-\sum_ip_{x,y}\log_2p_{x,y}
\end{equation}In this expression, $p_{x,y}$ is the joint probability of obtaining $x$ in a measurement of $X$, and $y$ in a measurement of $Y$.

The content of information which belongs to both, $X$ and $Y$, is called {\em mutual information}, $S_{X:Y}$, defined by:

\begin{equation}\label{eq26}
S_{X:Y}=S_X+S_Y-S_{X,Y}=S_X-S_{X|Y}
\end{equation}For classical states the above equality is always valid, but for quantum correlated states it is not! For two-qubits an entangled state, for instance, $S_{X,Y}=0$, whereas $S_X=S_Y=1$. On the other hand, $S_{X|Y}=0$. The difference between the classical and quantum mutual information is called {\em discord} \cite{ollivier}. A thorough review about discord and NMR is made in \cite{maziero2}. For a two-qubit system, $\rho_{AB}$, in a Bell diagonal state, Luo S. \cite{luo} found a simple analytical expression for this entropic-based quantum discord given by
\begin{eqnarray}
D(\rho_{AB}) = 2 + \sum_{k=0}^3\lambda_k\log_2\lambda_k - \frac{1-c}{2}\log_2(1-c)-\frac{1+c}{2}\log_2(1+c),
\end{eqnarray}
where $\lambda_k$ is the k-th eigenvalue of $\rho_{AB}$ and $c=\max\{|c_1|,|c_2|,|c_3|\}$. 

However, as this entropic formulation depends on numerical extremizations, analytical expressions are known for only few classes of states. For that reason, and based on which were previously defined for entanglement measurements, discord quantifiers based on geometric arguments were proposed \cite{modi}. By definition, entanglement captures the non-separability degree of a global state $\rho$. An entanglement geometric quantifier is calculated through the distance between a state and its closest separable state, $\sigma$, which can be written as a convex combination of product states,
\begin{eqnarray}
\sigma = \sum_{ij}p_{ij}\rho^A_i\otimes\rho^B_j
\end{eqnarray}
Analogously, a discord geometric quantifier is defined based on the distance between a state and its closest classical state. This classicality is associated with the application of a local projective measurement, then, in the two-qubit system example, this measurement can be applied on a subsystem only (asymmetric discord-type measures) or on both of them (symmetric discord-type measures). For an asymmetric discord-type geometric measure the distance is calculated to the closest classical-quantum state, defined as
\begin{eqnarray}
\chi = \sum_{i}p_i\ket{i}\bra{i}\otimes\sigma^B_i
\end{eqnarray}
The symmetric version, however, requires a classical-classical state
\begin{eqnarray}
\chi = \sum_{i,j}p_{i,j}\ket{i}\bra{i}^A\otimes\ket{j}\bra{j}^B
\end{eqnarray}
 
To calculate this distance it is necessary to choose a metric, that is why a whole set of geometric-based discord quantifiers can be found in the literature nowadays. However, only some of them are proved to be {\it bona fide} \cite{cianciaruso}.       

\subsubsection{NMR Discord and Relaxation Effects}

\begin{figure}
\begin{center}
\includegraphics[scale=0.6]{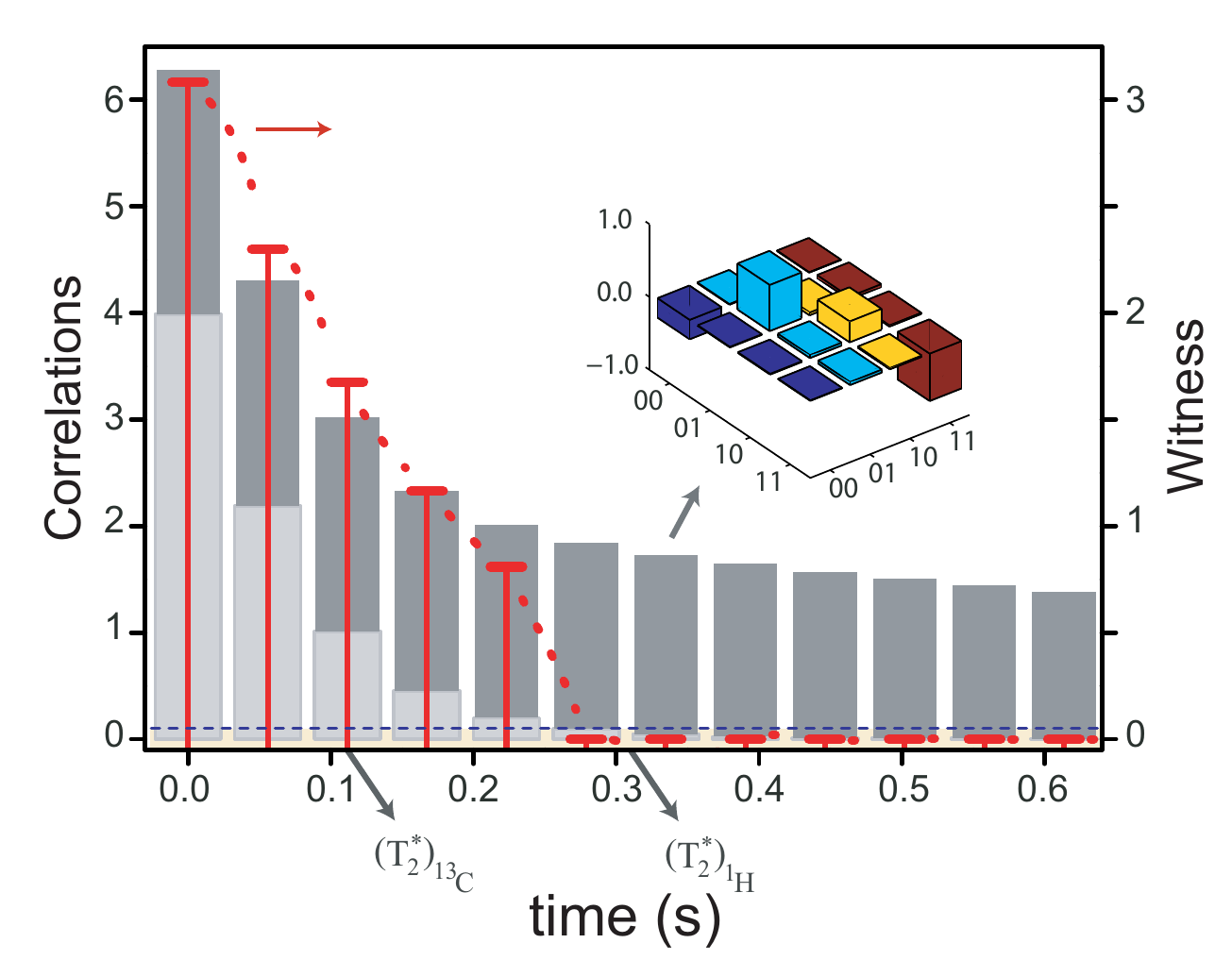}
\caption{The panel displays a witness and computed correlations of a quantum correlated state ($c_1 = 2\varepsilon$, $c_2 = 2\varepsilon$ and $c_3 = -2\varepsilon$) relaxed during a time interval, $t_{n}=n \delta t$ ($ \delta t = 55.7$ ms, $n=0,2,...,11$). The red tick bars represent the witness expectation value (proposed in Ref.~\cite{ruben2}), the grey bars display the quantum mutual information (total correlation), the dark grey section represents the amount of classical correlation, and the light grey section represents the quantum discord. In each experimental run, the correlations quantifiers were computed from the tomography data while the witness was directly measured \cite{ruben2, silva}. The classicality bound is represented by the blue doted line. The inset image shows the real part of the deviation matrix elements reconstructed by QST for an intermediate classically correlated state. The effective transversal relaxation times, shown below the figure, are $T_{2}^{\ast}=0.31$ s and $T_{2}^{\ast}=0.12$ s, for $^{1}$H and $^{13}$C nuclei, respectively. The correlations are displayed in units of ($ \varepsilon^{2} / \ln2$)bit. \newline  \centerline{Taken from Ref.~\cite{ruben2}.}}
\label{fig4_witness}
\end{center}
\end{figure}

Discord is an extremely useful quantity to detect quantum correlation. Fig. \ref{fig4_witness}, taken from Ref. \cite{ruben2} shows the time evolution of classical and quantum correlations of a two-qubit system under natural decoherence of a carefully prepared initial state. The picture shows the respective transverse relaxation times of the two-qubit system.

\begin{figure}
\begin{center}
\includegraphics[scale=0.7]{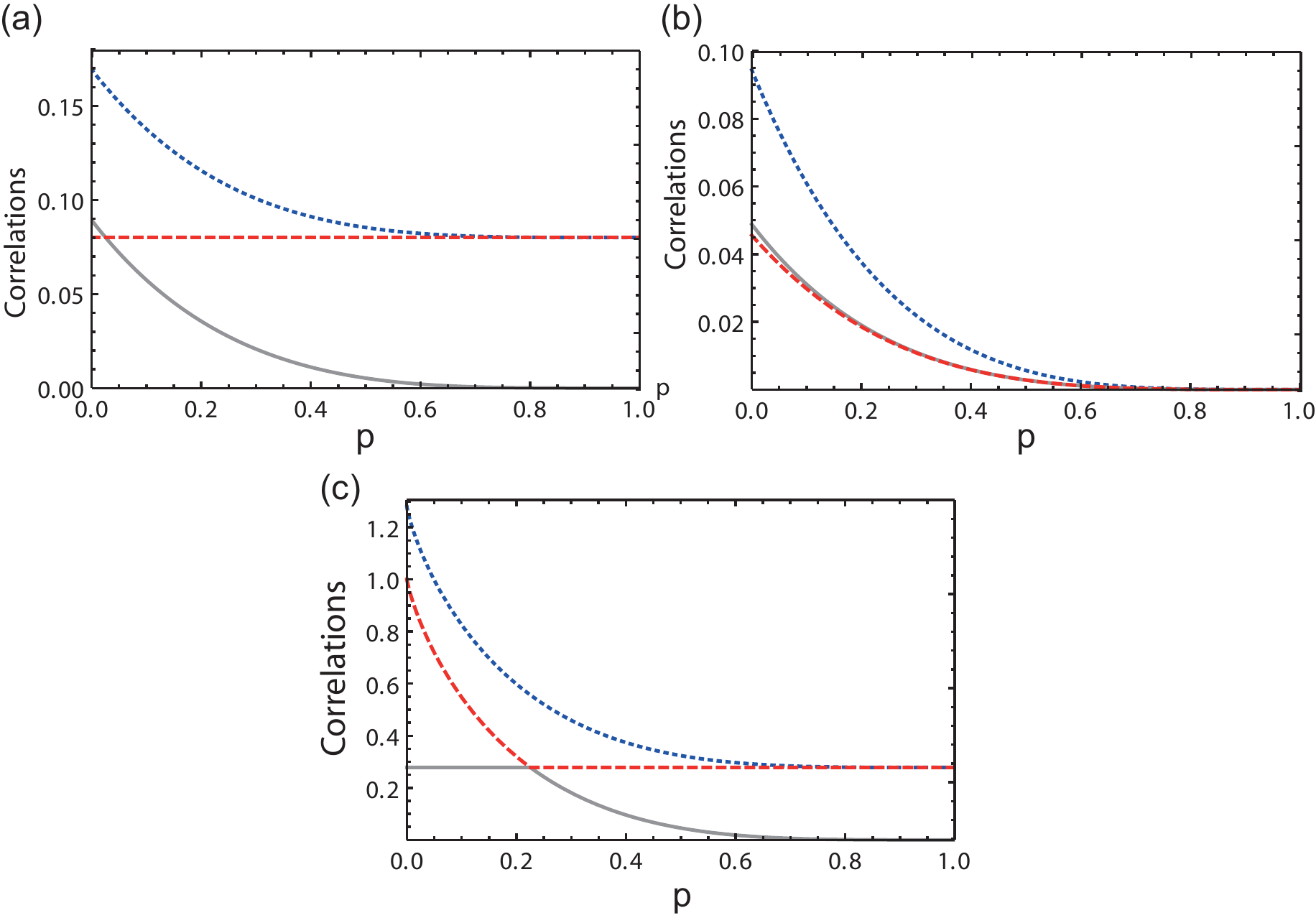}
\caption{Total (blue dotted line), classical (red dashed line), and quantum discord (gray continuous line) correlations for a Bell diagonal state evolving under local independent PD channels. In (a) the correlation triple is given by $\vec{c} = \{0.06, 0.30, 0.33\}$. In this case the classical correlation is not affected by the environment while the quantum correlation decays monotonically. In (b) $\vec{c} = \{0.25, 0.25, 0.00\}$ and all correlations decay monotonically. In (c) $\vec{c} = \{1.00, -0.60, 0.60\}$. For this state a sudden change occurs at $p_{SC}\approx 0.22$ and the quantum discord (classical correlation) remains constant (decays monotonically) for $p\leq 0.22$ with the opposite scenario taking place for $p\geq 0.22$. \newline  \centerline{Taken from Ref.~\cite{maziero2}.}}
\label{PDdecoh}
\end{center}
\end{figure}

In Ref. \cite{maziero} three general types of dynamics were identified for this two-qubit system under natural decoherence (phase flip, bit flip and bit-phase flip channels). These three categories depend on the relation between the correlation matrix elements. For the NMR natural phase flip decoherence, the three different types of dynamics are observed for (i) $|c_3|\geq|c_1|,|c_2|$, (ii) $|c_3| = 0$ and, the most interesting case, (iii) $|c_1|\geq|c_2|,|c_3|$ or $|c_2|\geq|c_1|,|c_3|$ and $|c_3| \neq 0$. Figure \ref{PDdecoh} (taken from Ref. \cite{maziero2}) shows these three dynamical possibilities for classical, quantum correlations and mutual information.

As pointed out in Ref. \cite{maziero}, the case (iii) reveals a peculiar sudden change in classical and quantum correlations. Afterwards, in Ref. \cite{mazzola} it was discovered that for a special class of initial states, quantum correlations are not destroyed by decoherence for times $t<\bar{t}$, while classical correlations decay. Then, for $t>\bar{t}$, classical correlations remain constant in time and quantum correlations are destroyed. This phenomenon was called Quantum Correlation Freezing. The classical correlation freezing is associated to the appearance of a pointer basis \cite{cornelio}. 

Ref.\cite{silva} shows the first experimental observation of quantum correlations freezing phenomenon, where a direct measurement procedure was applied. Until then, the NMR experiments were recorded applying full quantum state tomography. This work proved that, since discord quantifiers for Bell diagonal states depend only on the correlation function elements and these elements are proportional to the NMR signal (when a proper set of unitary rotations are applied), it is possible to avoid the expensive quantum state tomography and perform a direct measurement. Mathematically, it means
\begin{eqnarray}
c_i = \textrm{Tr}\{(\sigma_i\otimes\sigma_i)\rho\} = \textrm{Tr}\{(\sigma_i\otimes I)\xi_i\}, \\
\xi_i = U_i\rho U_i^{\dagger},
\end{eqnarray}
where $U_i$ are the properly unitary rotations as described in Ref.\cite{silva} and $\textrm{Tr}\{(\sigma_i\otimes I)\xi_i\}$ is the detected NMR signal, as described in Eq. (\ref{eq18}).

\begin{figure}
\begin{center}
\includegraphics[scale=0.7]{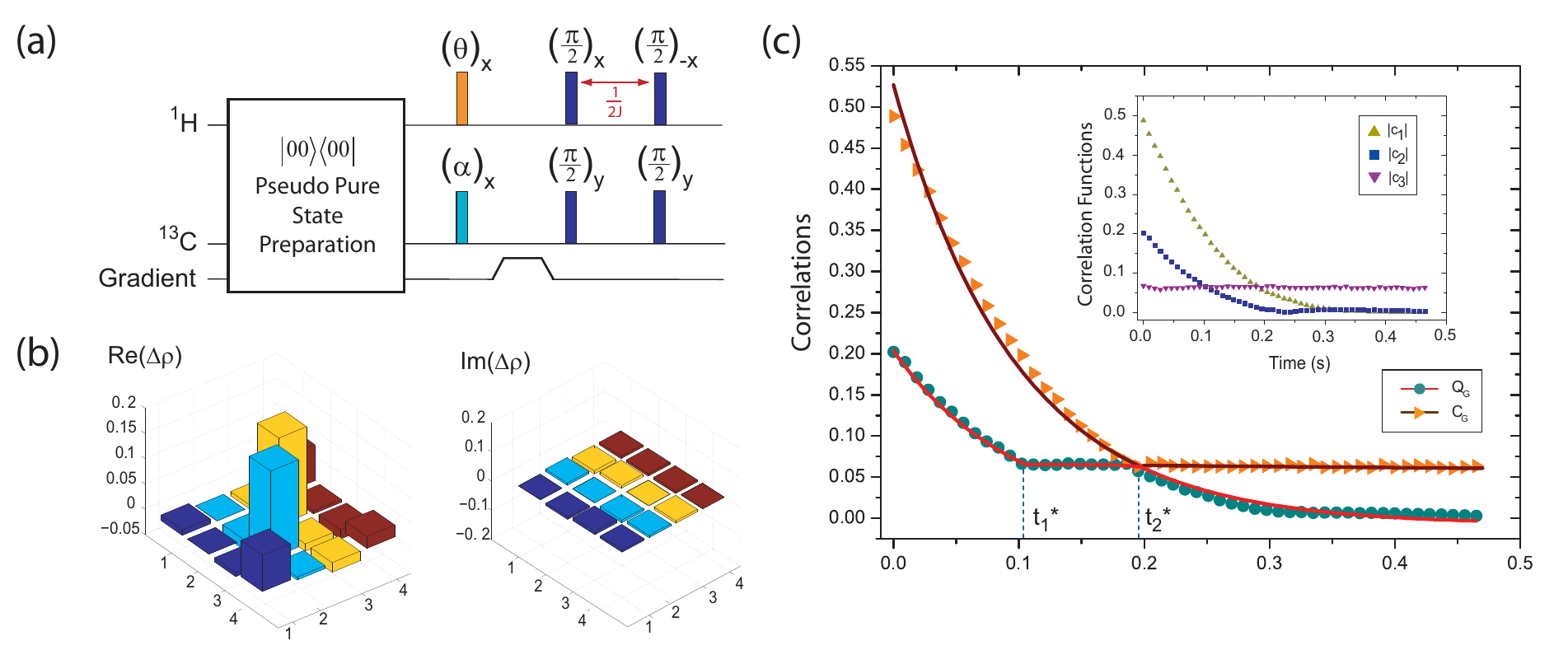}
\caption{(a) Schematic representation of the pulse sequence employed to obtain a deviation matrix in the form of a Bell diagonal state. (b) Experimentally reconstructed block diagrams for real and imaginary parts of the deviation matrix related to the Bell diagonal initial state with $c_1 = 0.49$, $c_2 = 0.20$ and $c_3 = 0.067$. The curves in (c) denote the time evolutions of quantum ($Q_G$, bullet) and classical ($C_G$, triangle) correlations, respectively. The dots represents the experimental results and the solid lines are the theoretical predictions. In the inset we detail the time evolutions of $|c_1|$ (yellow upward triangles), $|c_2|$ (blue squares), and $|c_3|$ (purple downward triangles) experimentally obtained for the PD decoherence process.\newline \centerline{Taken from Ref.~\cite{paula}.}}
\label{DSC12}
\end{center}
\end{figure}

In the same line, Ref. \cite{paula} reported an NMR experiment of {\em double sudden change}, theoretically predicted in Ref. \cite{montealegre}, where two different two-qubit NMR setups were applied. The first one was performed on a Varian $500$ MHz spectrometer on a liquid state Carbon-$13$ enriched Chloroform sample ($^{13}$CHCl$_3$) at room temperature. In this case, the two-qubit are encoded in the $^1H$ and $^{13}C$ spin-1/2 nuclei. The measurement of characteristic relaxation times provided $T_1^C \approx 12.46$ s, $T_2^C \approx 0.15$ s, $T_1^H \approx 7.53$ s and $T_2^H \approx 0.27$ s. Since $T_1 >> T_2$ for both spins, and being the experiment evaluation time smaller than $0.5$ s, the GAD channel effects can be neglected and the entire relaxation mechanism can be described effectively by a PD channel only. The prepared initial state corresponds to a Bell state with $\vec{c} = \{0.49, 0.20, 0.067\}$, which satisfies condition (iii). In Fig. \ref{DSC12} it is possible to observe clearly one sudden-change in classical correlations ($C_G$) and two points of sudden change in quantum correlation dynamics ($Q_G$). 

\begin{figure}
\begin{center}
\includegraphics[scale=0.4]{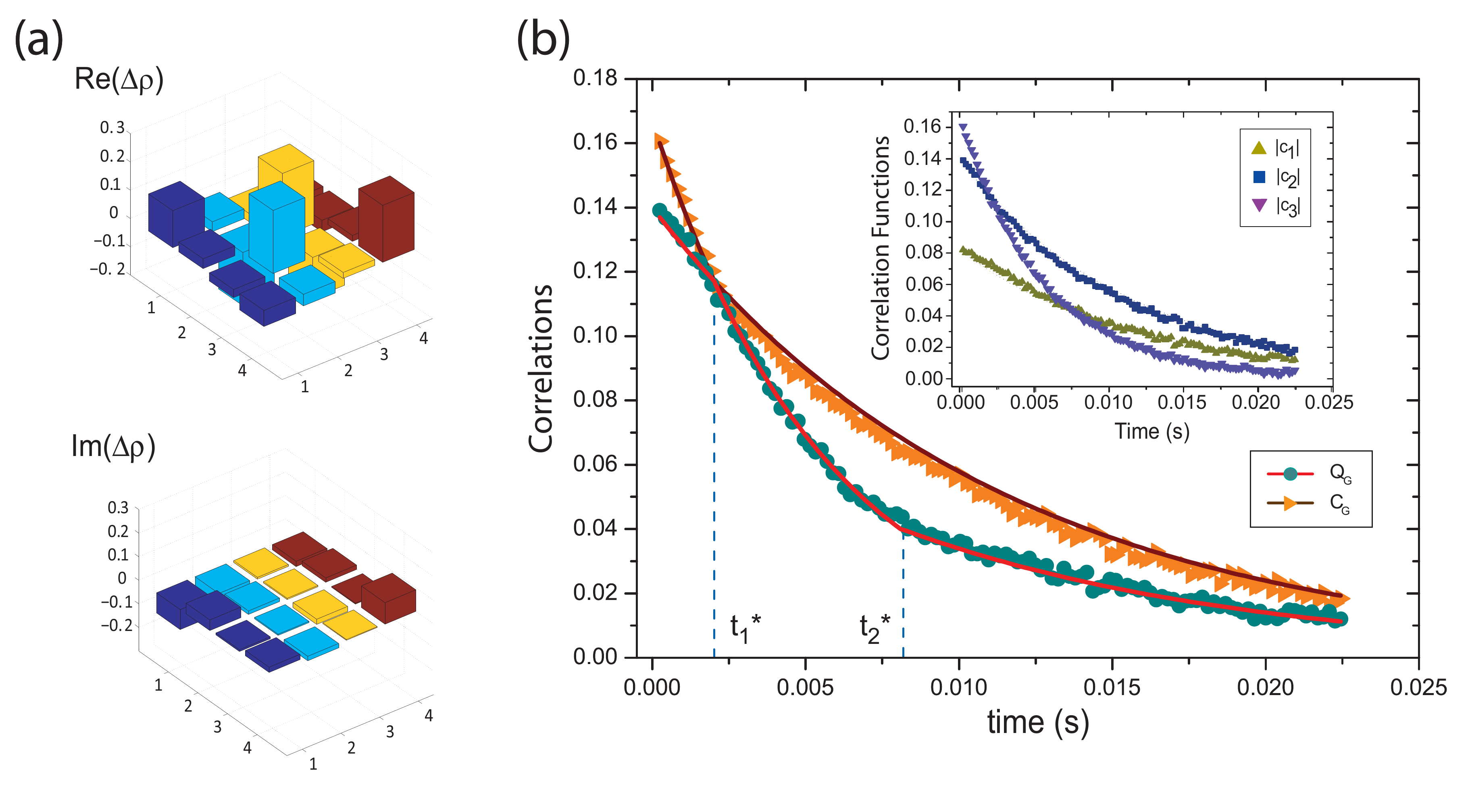}
\caption{(a) Experimentally reconstructed block diagrams for real and imaginary parts of the deviation matrix related to the Bell diagonal initial state with $c_1 = 0.08$, $c_2 = 0.14$ and $c_3 = 0.16$. The curves in (b) denote the time evolutions of quantum ($Q_G$, bullet) and classical ($C_G$, triangle) correlations, respectively. The dots represents the experimental results and the solid lines are the theoretical predictions. In the inset we detail the time evolutions of $|c_1|$ (yellow upward triangles), $|c_2|$ (blue squares), and $|c_3|$ (purple downward triangles) experimentally obtained for the GAD decoherence process.\newline \centerline{Taken from Ref.~\cite{paula}.}}
\label{DSC32}
\end{center}
\end{figure}

The other system was a spin-$3/2$ NMR quadrupolar setup encoded in a liquid crystal sample with an NMR detectable sodium nuclei (more details about this setup is found in Ref. \cite{lawson, radley}). The experiment was performed in a Varian $400$ MHz spectrometer at room temperature. Considering that in this case a two-qubit system is encoded in a single nuclei (which in the presence of a strong static magnetic field is described by four energy levels), the relaxation is described by the GAD channel and a modified PD channel called Global Phase Damping (GPD), which acts on both logical qubits simultaneously, and was proposed in Ref. \cite{souza}. However, GPD does not act on the cross-diagonal terms of the deviation matrix, meaning that Bell diagonal states are not affected by them and the decoherence is completely dictated by GAD channel. The initial Bell diagonal state prepared in this case corresponds to $\vec{c} = \{0.08, 0.14, 0.16\}$ and was implemented applying a strongly modulated pulse (SMP) method, where the radio frequency pulses are numerically optimized, as described in Ref. \cite{fortunato}. Althought the relaxation is described by a non-conservative energy channel, it is also possible to observe a single and a double sudden change on classical and quantum correlations, respectively, as shown in Fig. \ref{DSC32}. In this case classical correlations do not remain constant, so a pointer basis is not reached. And despite the quantum correlation changes its relaxation rate, it is destroyed during the entire process, while in the other system (under PD action) it is frozen between $t_1^\star$ and $t_2^\star$. 

\subsubsection{NMR Observation of Freezing Phenomenon}
The frozen quantum correlation phenomenon was theoretically explained in detail in Ref. \cite{cianciaruso}, where it was demonstrated that for an initial Bell diagonal state (two-qubit system) satisfying
\begin{eqnarray}
c_1(0) = \pm 1, \, c_2(0) = \mp c_3(0),
\end{eqnarray}
under a PD channel effect, for any reliable geometric quantum coherence quantifier, the freezing phenomenon will be observed from $t = 0$ (initial time) until $t = t^{\star}$,
\begin{eqnarray}
t^* = -\frac{1}{2\gamma}\ln\frac{|c_3(0)|}{|c_1(0)|},
\end{eqnarray}
after that the dynamical evolution occurs in an exponential way, and the relaxation rate is different for each adopted quantifier.

\begin{figure}
\begin{center}
\includegraphics[scale=0.5]{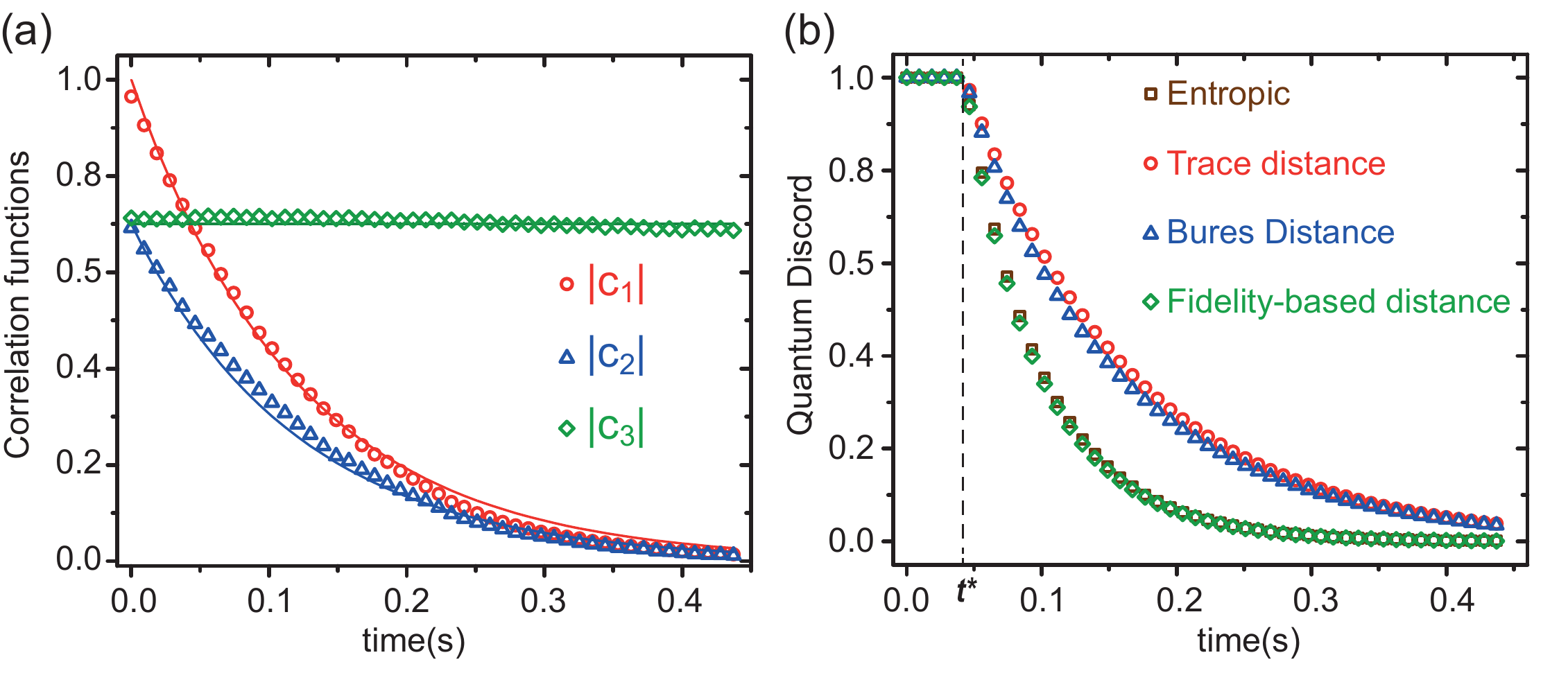}
\caption{(a) Time evolution of correlation function elements: $|c_1|$ (circle), $|c_2|$ (triangle), $|c_3|$ (diamond). (b) Time evolution of quantum discord quantified by entropic (square), trace distance (circle), Bures distance (triangle) and fidelity based distance (diamond) measures. The freezing phenomenon is observed for all quantifiers from $t=0$ to $t=t^\star$, when an exponential evolution starts differently for each one \cite{silva}.}
\label{universal_discord}
\end{center}
\end{figure}

Fig. \ref{universal_discord} shows those experimental results, where the $^{13}$CHCl$_3$ sample was once more employed as a two-qubit system setup. A pulse sequence analogous to Fig. \ref{DSC12}-a was applied, choosing $\theta$ and $\alpha$ appropriately to produce the initial state $c_1 = 1$, $c_2 = 0.7$ and $c_3 = -0.7$. Then, we observe that quantum discord calculated from entropic discord, trace, Bures and fidelity-based distances, is frozen until $t^{\star} = 0.04$ s.

For more general systems, $N>2$ qubits, the quantifier called Global Quantum Discord, defined in Ref. \cite{xu, rulli}, predicts that frozen quantum correlations will be observed for systems with an even number of qubits, while it will never be observed for the odd case.   

\begin{figure}
\begin{center}
\includegraphics[scale=0.5]{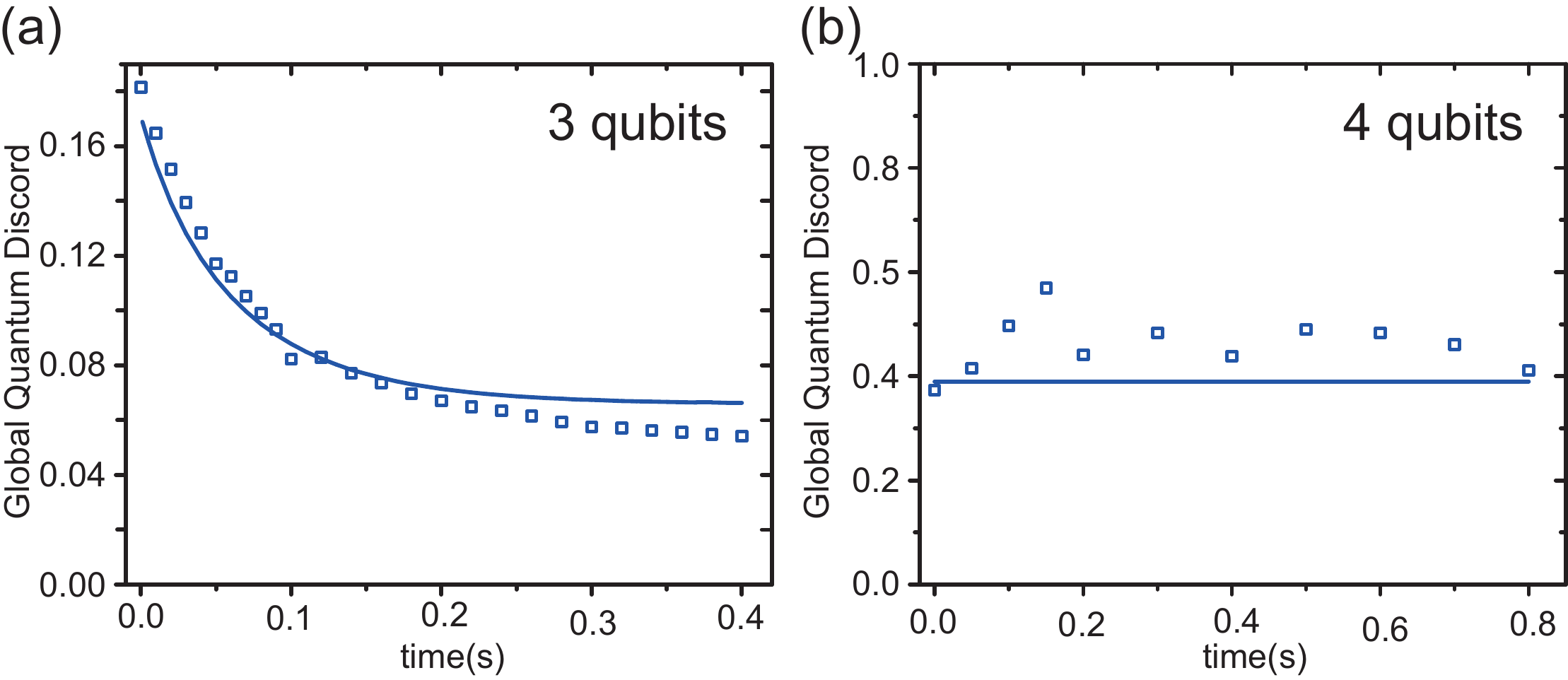}
\caption{\label{GQD34qb} Global Quantum Discord measured for NMR setups of (a) $3$ and (b) $4$ qubits. The dots correspond to experimental data and lines to theoretical predictions.}
\end{center}
\end{figure}

\begin{figure}
\begin{center}
\includegraphics[scale=0.4]{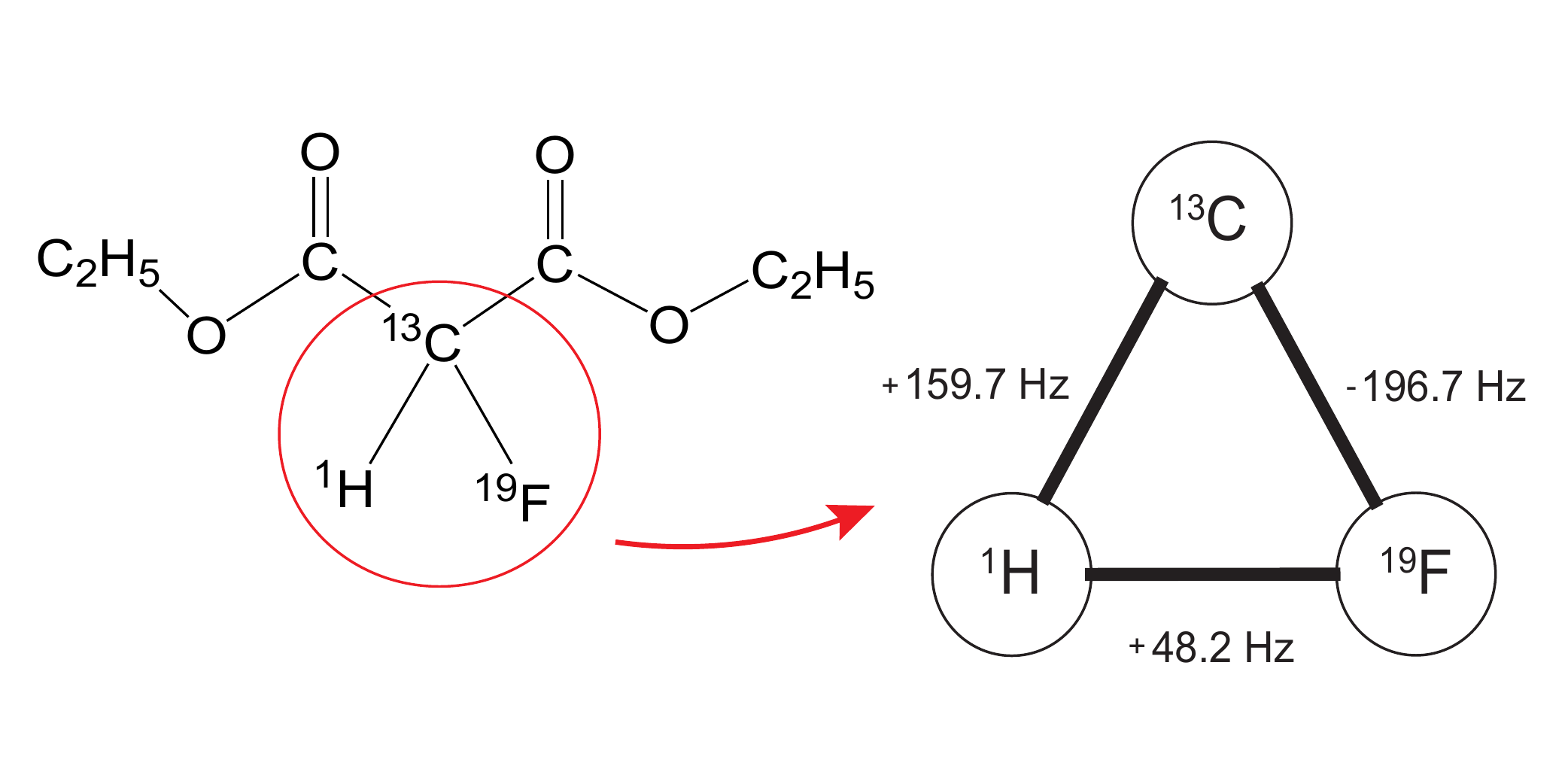}
\caption{\label{molecule3} Structural formula of Diethyl 2-Fluoromalonate-2-$^{13}$C. The three-qubit were encoded in $^1$H, $^{19}$F and $^{13}$C. The coupling topology for this three-qubit system is shown.}
\end{center}
\end{figure}

In order to experimentally observe those predictions, see Fig.~\ref{GQD34qb}, a three-qubit system was encoded on a sample of Diethyl 2-Fluoromalonate-2-$^{13}$C dissolved in CDCl$_3$, for which the structural formula and coupling topology is shown in Fig.~\ref{molecule3}. The three-qubit were encoded in the $^1$H, $^{19}$F and $^{13}$C nuclear spins. In this molecule each qubit is coupled to the two other ones. All scalar coupling constants were measured in E-COSY (exclusive correlation spectroscopy) type experiments: $J_{HF} \approx +48.2$ Hz, $J_{HC} \approx +159.7$ Hz and $J_{FC} \approx -196.7$ Hz. This experiment was performed in a Bruker AVIII $600$ MHz spectrometer, equipped with a QXI $600$ MHz S3 five channels ($^2$H, $^1$H, $^{13}$C, $^{15}$N, $^{19}$F) probe with z-gradient, at room temperature.      

\begin{figure}
\begin{center}
\includegraphics[scale=0.5]{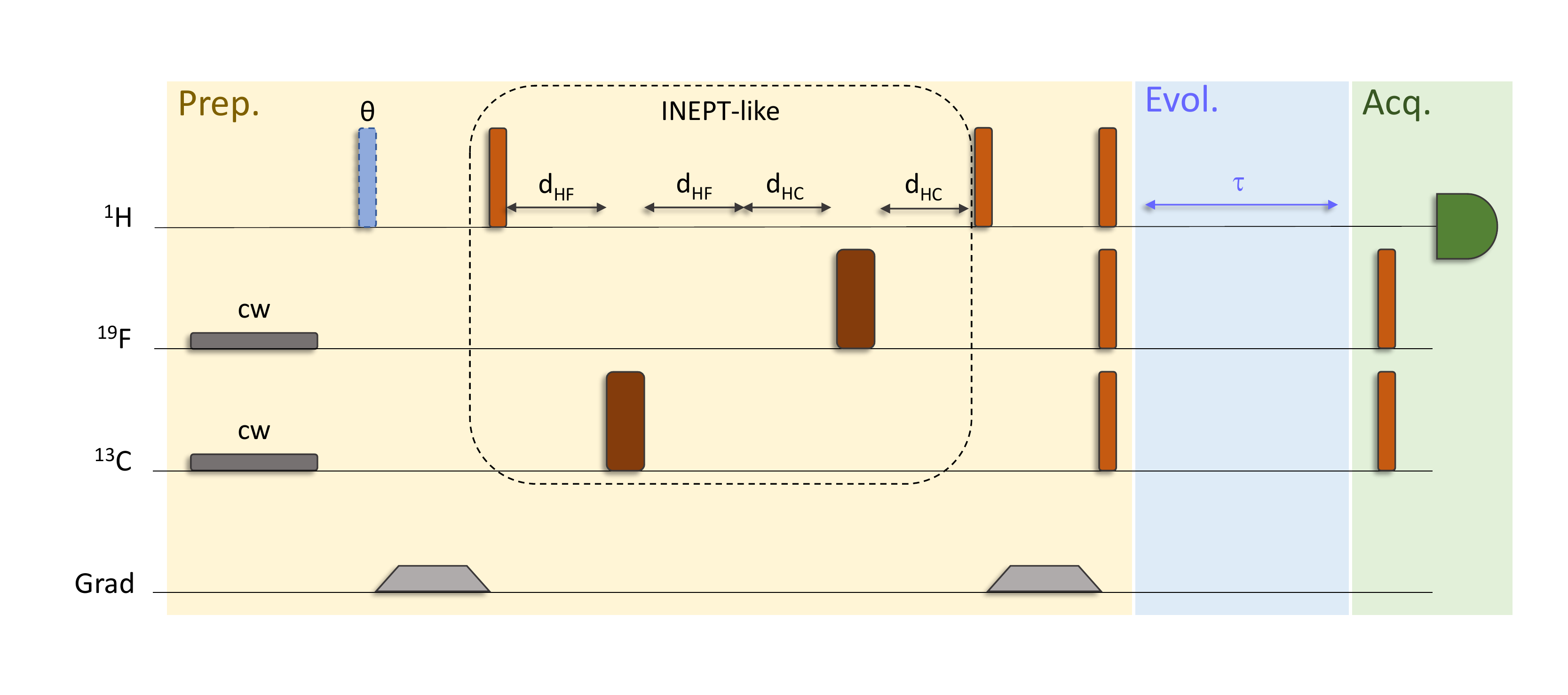}
\caption{\label{pseq3Q} Pulse sequence for the three-qubit experiment. First, a continuous wave (cw) pulse is applied to the $^{19}$F and $^{13}$C spins in order to avoid contributions from those nuclei. Then, a pulse with variable length (angle) is implemented to produce the correct scaling in the correlation function elements.  Next, an INEPT-like sequence (where thin bars represent $\pi/2$ pulses and large bars $\pi$ pulses) is applied to produce the desired multispin term (like $I^H_zI^F_zI^C_z$), where $d_{kl} = 1/(4J_{kl})$. After that, as at this point the multispin term is aligned along the longitudinal axis, a gradient pulse is applied to eliminate other terms and guarantee the state quality. Then, $\pi/2$ pulses are applied, with the appropriate phase, to produce the $M^3_3$ state term. The system is allowed to evolve freely and later $\pi/2$ pulses, with appropriate phases, are applied to produce an NMR detectable (single quantum) term in the $^1$H channel. }
\end{center}
\end{figure}

The initial $M^3_3$ state, 
\begin{eqnarray}
\rho = \frac{1}{8}\mathbb{I}^{\otimes 3} + c_1\sigma_x^{\otimes 3} + c_2\sigma_y^{\otimes 3} + c_3\sigma_z^{\otimes 3},
\end{eqnarray}
was prepared applying the pulse sequence shown in Fig.~\ref{pseq3Q}. To distinguish the physical qubits, we will associate each $\sigma_i^{\otimes 3}$ term to the spin operators $I_i^HI_i^FI_i^C$. Each of these terms is independent of the others and also interacts independently with the environment (considering the same kind of decoherence process described before for the two-qubit system), therefore each term was prepared separately. First, a continuous wave (cw) pulse is applied to $^{19}$F and $^{13}$C to guarantee that all terms will be generated from $^1H$ magnetization. Then, a pulse with an appropriate angle, $\theta = \{0.8, 1.27, 1.27\}$ rad, is applied followed by a gradient pulse, which guarantees that only $\cos\theta$ of initial magnetization will survive, producing the correct scaling $\vec{c} = \{0.7, 0.3, 0.3\}$, after normalizing with full magnetization spectra. Then, an INEPT-like sequence produces the multispin term ($I_z^HI_z^FI_z^C$) from $^1H$, and a gradient pulse guarantees the initial state quality. The $\pi/2$ pulses applied on all nuclei, with appropriate phases, produce each density matrix term and are followed by a free evolution period. Finally, each element of the coherence triple is directly measured  by producing single-quantum terms (like $I_x^HI_z^FI_z^C$), similarly to what was implemented in Ref.\cite{silva}, where it was shown that a direct measure is as good as a full state tomography. 

\begin{figure}
\begin{center}
\includegraphics[scale=0.7]{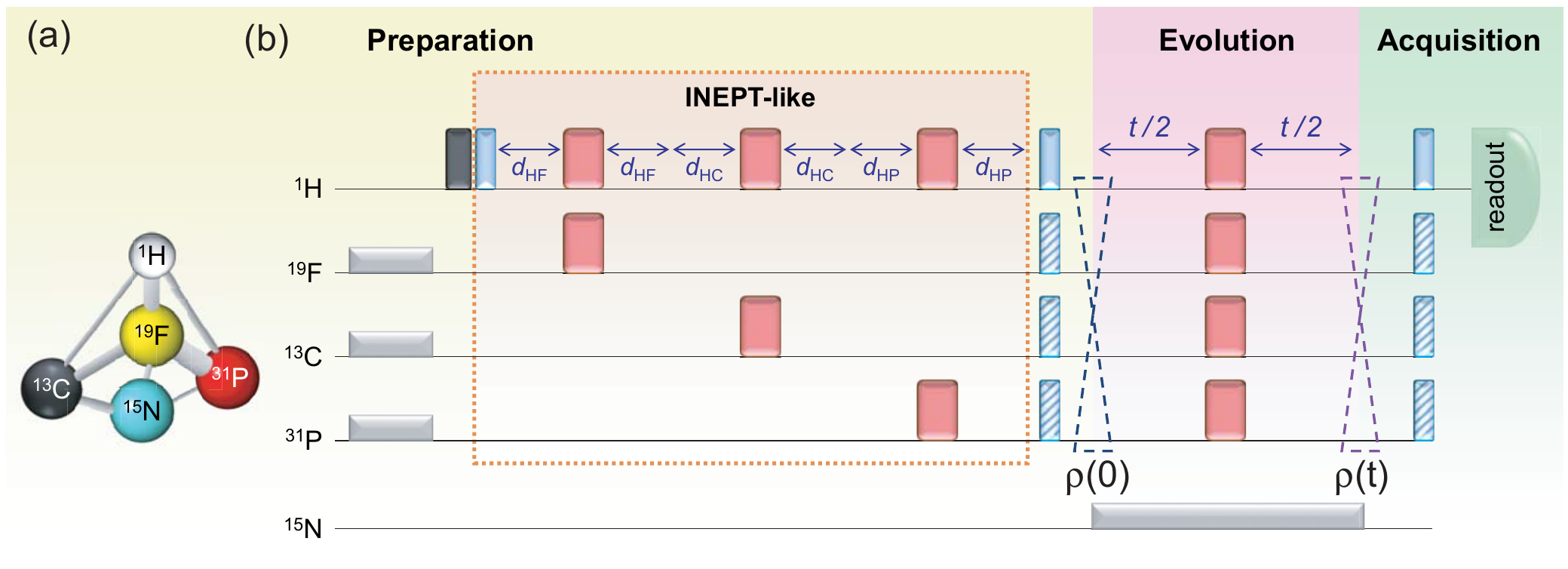}
\caption{\label{pseq4Q} Pulse sequence (with time flowing from left to right) to prepare four-qubit generalised Bell states encoded in the $^1$H, $^{19}$F, $^{13}$C, and $^{31}$P nuclear spins of the $^{13}$C$^O$-$^{15}$N-diethyl-(dimethylcarbamoyl)fluoromethyl-phosphonate molecule (whose coupling topology is illustrated in (a) by an INEPT-like procedure, where $d_{kl} = 1=(4J_{kl})$ and $J_{kl}$ is the scalar coupling between spins $k$ and $l$. Light-gray rectangles denote continuous-wave pulses, used to decouple the $^{15}$N nucleus. The dark grey bar denotes a variable pulse, applied to set the desired correlation triple $\{c_j(0)\}$. Thicker (red) and thinner (blue) bars denote $\pi$ and $\pi/2$ pulses, respectively; the phases of the striped $\pi/2$ pulses were cycled to construct each density matrix element. After the preparation stage, the system
was left to decohere in its environment; $\pi$ pulses were applied in the middle of the evolution to avoid $J_{kl}$ oscillations. The final $\pi/2$ pulses served to produce a detectable NMR signal in the $^1$H spin channel. \newline \centerline{Taken from Ref.~\cite{silva2}.}}
\end{center}
\end{figure}

The even case was observed in a four-qubit system encoded in the $^{13}$C$^O$-$^{15}$N-diethyl-(dimethylcarbamoyl)fluoromethyl-phosphonate compound, whose coupling topology is shown in Fig.~\ref{pseq4Q}; a detailed description of its synthesis can be found in Ref. \cite{marx}. This molecule contains five NMR-active spins ($^1$H, $^{19}$F, $^{13}$C, $^{31}$P and $^{15}$N), so a continuous wave (cw) pulse was applied to decouple $^{15}$N. The initial ${\cal M}^3_4$ (alias generalised BD) state,
\begin{equation}
\rho(0) = \frac{1}{16}\left(\mathbb{I}^{\otimes 4} + c_1(0)\sigma_1^{\otimes 4} + c_2(0)\sigma_2^{\otimes 4} + c_3(0)\sigma_3^{\otimes 4}\right)
,
\end{equation}
was prepared according to the pulse sequence displayed in Fig.~\ref{pseq4Q}. Similarly to what was implemented for the three-qubit case, each term of these density matrix was prepared separately. 

Those phenomena were theoretically and experimentally discussed for quantum coherence in  Refs.~\cite{bromley, silva2}.

\subsubsection{Discord and the interferometric power of quantum states}

Liquid state NMR was also employed to study quantum correlations in a quantum metrology scenario \cite{girolami}.
Unlike most common metrological applications, the object of investigation was the role of quantum discord in an interferometric 
configuration when only the spectrum of the generating Hamiltonian is known. In more details, we consider a 
bipartite probe state $\rho_{AB}$ entering a two-arm channel, in which only the subsystem $A$ is affected by a local
unitary $U_{A} = e^{-i\varphi H_A}\otimes\mathbb{I}$, where $\varphi$ is the parameter to be estimated and $H_A$ is the 
generating Hamiltonian. The information about $\varphi$ is obtained through the application of an estimator $\tilde\varphi$
over the output state $\rho_{AB}^{\varphi} = U_{A}\rho_{AB}U_{A}^{\dagger}$. If we have full knowledge of $H_A$, the maximum 
achievable precision is obtained when the input state has maximum coherence in the eigenbasis of $H_A$, and no 
quantum correlation between subsystems $A$ and $B$ is necessary \cite{MAlmeida}. A figure of merit to quantify the
information about $\varphi$ encoded in $\rho_{AB}^{\varphi}$ is the quantum Fisher information \cite{Braunstein94}. The quantum
Cramer-Rao bound sets a lower limit to the variance of $\tilde\varphi$, the estimated parameter, as 
$\text{Var}_{\rho_{AB}^{\varphi}}(\tilde\varphi) \geq [\nu F(rho_{AB}, H_A)]^{-1}$, where $\nu$ is the number of repetitions of 
the experiment with identical copies  and $F(\rho_{AB}, H_A)$ is the quantum Fisher information \cite{Helstrom76}.

When a "black box" paradigm is considered, i.e., if only the spectrum of $H_A$ is known a priori, some aspects 
of game are changed. One can imagine this setting as as a referee deciding the local transformation over $A$, after 
the probe state $\rho_{AB}$ is prepared. After the application of $U_A$, the referee discloses his choice for $H_A$, 
allowing the experimenter to perform the respective optimal estimator. For the worst case scenario, the precision is minimal 
over all $H_A$ and can be computed using the quantum Fisher information as
\begin{eqnarray}\label{defIP}
P^{A}(\rho_{AB}) = \frac{1}{4}\inf_{H_A} F(\rho_{AB}, H_A),
\end{eqnarray}
\noindent where the infimum is over all Hamiltonians with a given spectrum and the normalization factor $1/4$ is 
solely for convenience. $P^{A}(\rho_{AB})$, termed interferometric power of $\rho_{AB}$, is a well defined measure of quantum correlations of a bipartite state, as shown
in \cite{girolami} and discussed in details by Bogaert and Girolami \cite{Davidechapter}. On one hand, if $\rho_{AB}$
is not discordant the interferometric power vanishes, since there is a $H_A$ such that $[\rho_{AB}, H_A] = 0$, and no information
about $\varphi$ can be retrieved. On the other hand, the degree of quantum correlations of $\rho_{AB}$ not only guarantees 
a minimal precision but also quantifies the usefulness of $\rho_{AB}$ as a resource to estimate $\varphi$. 

In the particular case of the subsystem $A$ being a qubit, a computable and closed formula was obtained 
for the interferometric power \cite{girolami}
\begin{eqnarray}\label{closedIP}
P^{A}(\rho_{AB}) = \zeta_{\min}[M],
\end{eqnarray}
\par\noindent where $\zeta_{\min}[M]$ is the minimal eigenvalue of the 3 X 3 matrix $M$:
\begin{eqnarray}\label{MMM}
M_{mn} = \frac{1}{2}\sum_{i,l; q_i + q_l \neq 0} \frac{(q_i - q_l)^2}{q_i + q_l}
\langle\psi_i\vert\sigma_{mA}\otimes\mathbb{I}_B\vert\psi_l\rangle
\langle\psi_l\vert\mathbb{I}_A\otimes\sigma_{nB}\vert\psi_l\rangle,
\end{eqnarray}
\par\noindent and $\{ q_i, \vert\psi_i\rangle\}$ is the set of eigenvalues and eigenvectors of $\rho_{AB}$, respectively.

The role of interferometric power was observed experimentally, in a proof-of-principle implementation, with the 
quantum state $\rho_{AB}$ encoded in the nuclear spins of a $^{13}$C-labelled chloroform sample. Two different classes of 
states were compared, a discordant and a classical-quantum ones. The chosen families are
\begin{eqnarray}\label{statesIP}
\rho_{AB}^{Q} = \frac{1}{4}\left(
\begin{array}{cccc}
1+p^2 & 0 & 0 & 2p \\
0 & 1-p^2 & 0 & 0 \\
0 & 0 & 1-p^2 & 0 \\
2p & 0 & 0 & 1+p^2
\end{array}\right), \quad
\rho_{AB}^{C} = \frac{1}{4}\left(
\begin{array}{cccc}
1 & p^2 & p & p \\
p^2 & 1 & p & p \\
p & p & 1 & p^2 \\
p & p & p^2 & 1
\end{array}\right).
\end{eqnarray}

The parameter $p$ quantifies the purity of both classes, as ${\rm Tr}(\rho_{AB}^{(Q, C)})^2 = 1/4(1+p^2)^2$, and 
$0 \leq p \leq 1$. This allows a fair comparison between quantum and classical states, since for a given value of $p$ we have 
the same degree of mixedness. The states $\rho_{AB}^{C}$ are classically correlated, with $P^{A}(\rho_{AB}^{C}) = 0$ for any $p$,
while $\rho_{AB}^{Q}$ has discord monotonically increasing for $p > 0$.

\begin{figure}
\centering
\includegraphics[width=0.6\linewidth]{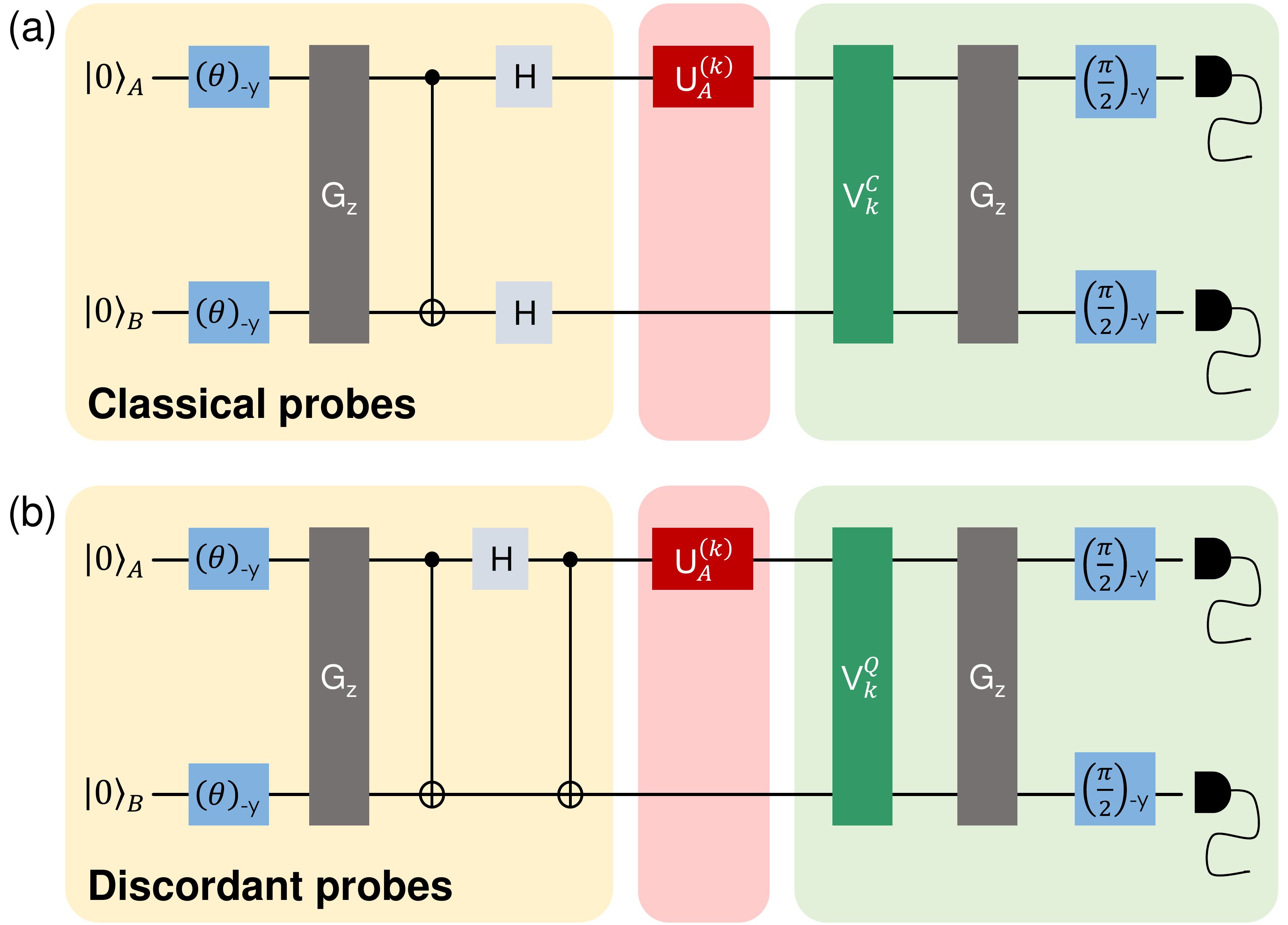}
\caption{Experimental scheme for black box parameter
estimation with NMR. (a) Classical probes. (b) Discordant probes. The protocol is divided in three steps:  probe
state preparation (yellow);  black box transformation (red);  optimal
measurement (green)}. \newline
\centerline{Taken from Ref. \cite{girolami}.}
\label{CircuitIP}
\end{figure}

The state preparation comprises three steps. First, the pseudopure state $\vert 00\rangle\langle 00\vert$ is prepared 
using a spatial average \cite{cory}. Secondly, the purity is chosen with $(\theta)_{-y}$ pulses on both nuclei, followed 
by a pulsed gradient field $G_z$. The final step consists in the application of the combination of CNOT and Hadamard gates
as depicted in Fig. \ref{CircuitIP} (a) for classical probes and Fig. \ref{CircuitIP} (b) for the quantum ones.

\begin{figure}
\centering
\includegraphics[width=0.6\linewidth]{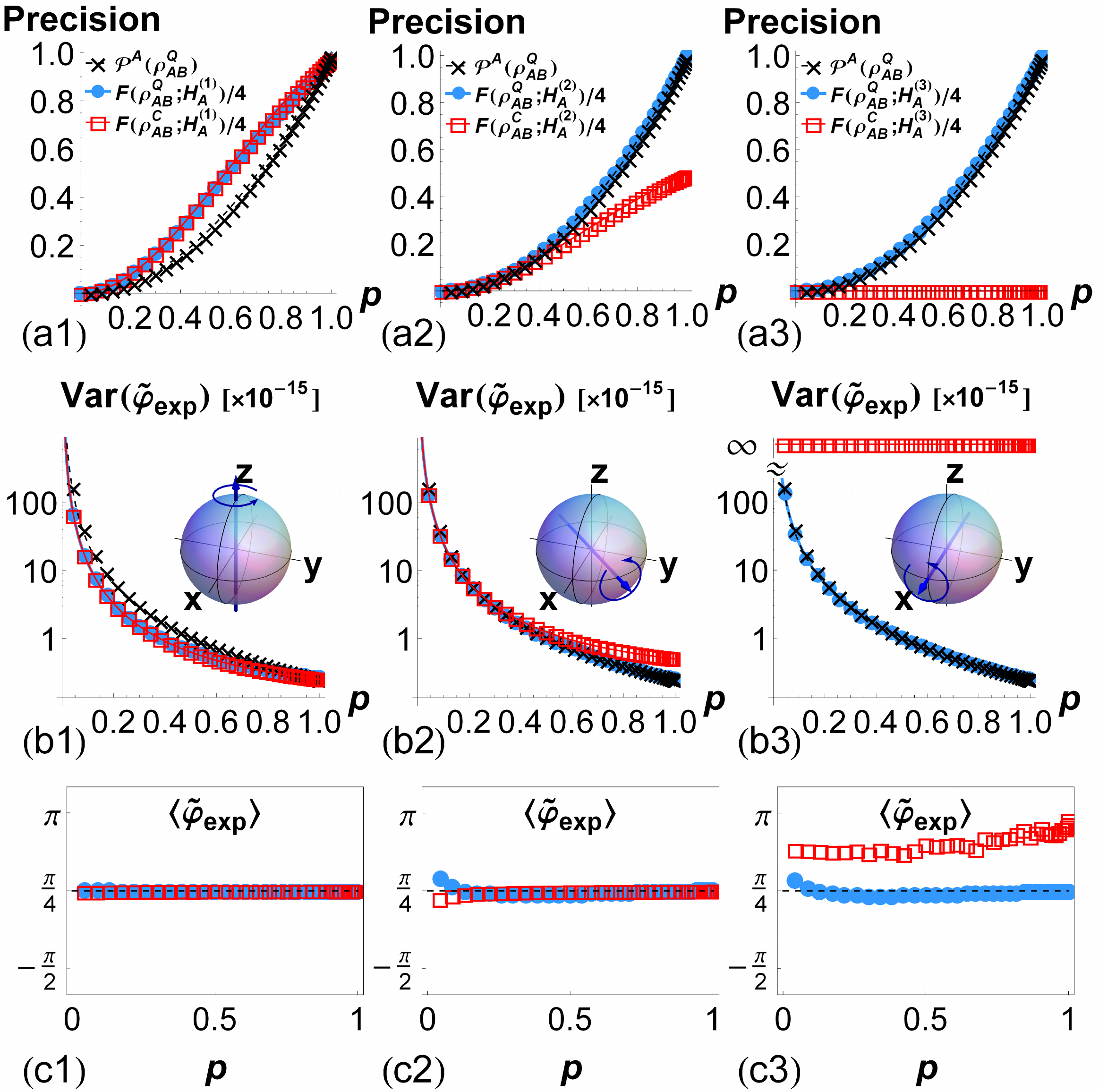}
\caption{Experimental results. Each column corresponds to a different black box 
setting $H^{(k)}_A$, $k = 1, 2, 3$, and the set directions are depicted in the insets of row (b).
Empty red squares refer to data from classical probes $\rho_{AB}^{C}$ and filled blue circles 
refer to data from discordant probes $\rho_{AB}^{Q}$. Both classes depend on the purity, quantified 
by the parameter $p$. The first row shows the measured quantum Fisher 
information normalized by a factor of $4$ for each setting, along with the lower bound provided by 
$P^{A}(\rho_{AB}^{Q})$. The middle row (b) presents the measured variances, together with a theoretical 
prediction for the saturation of the Cramer-Rao bound. Last, row (c) depicts the inferred mean value 
for each setting. The lines refer to theoretical predictions. \newline
\centerline{Taken from Ref. \cite{girolami}.}
}
\label{IPresults}
\end{figure}

To calculate the interferometric power, a full quantum state tomography was performed to each prepared state, with a mean fidelity
of $(99.7 \pm 0.2)$ with the theoretical density matrices of Eq. (\ref{statesIP}). The interferometric power was computed using 
the closed formula of Eq. (\ref{closedIP}) on the reconstructed states and is displayed in the first row of Fig. \ref{IPresults}, 
with an excellent agreement with the theoretical expectation, $P^{A}(\rho_{AB}^{Q}) = p^2$.

For each fixed probe, three choices for $H_A$ were performed. $H_{A}^{(1)} = \sigma_{z}^{A}\otimes\mathbb{I}_B$,
$H_{A}^{(2)} = (\sigma_{x}^{A} + \sigma_{y}^{A})/\sqrt{2}\otimes\mathbb{I}_B$ and $H_{A}^{(3)} = \sigma_{x}^{A}\otimes\mathbb{I}_B$.
These three Hamiltonians encompass the worst and best settings for $H_{A}^{(1)}$ and $H_{A}^{(3)}$, respectively, while
$H_{A}^{(2)}$ is an intermediate case \cite{girolami}. In all experiments, the parameter $\varphi$ was set to $\varphi_0 = \pi/4$.

Finally, the optimal measurement is carried out to estimate $\varphi$, for all six combinations of input states and black boxes. 
This set of estimators is given by the eigenvectors of the symmetric logarithmic derivative 
$L_{\varphi} = \sum l_j \vert\lambda_j \rangle\langle\lambda_j \vert$, which satisfies $\partial_{\varphi}\rho_{AB}^{\varphi} = 
\frac{1}{2}(\rho_{AB}^{\varphi}L_{\varphi} + L_{\varphi}\rho_{AB}^{\varphi})$. The quantum Fisher information is given as 
$F(\rho_{AB}, H_a) = {\rm Tr}(\rho_{AB}^{\varphi}L_{\varphi}^{2})$ $= 4 \sum_{i,l; q_i + q_l \neq 0} \frac{(q_i - q_l)^2}{q_i + q_l}
\vert\langle\psi_{i}\vert H_{A}\otimes\mathbb{I_B}\vert\psi_l\rangle\vert^2$, where $\{q_i,\vert\psi_i\rangle\}$ is the set of eigenvalues
and eigenvectors of $\rho_{AB}$ \cite{MGAParis}. The readout procedure comprises a global rotation into the eigenbasis 
of $L_{\varphi}$, shown as $V_{k}^{(C, Q)}$\footnote{All the $V_{k}^{(C,Q)}$ transformations, along their implementations,
are shown in \cite{girolami}.} in Fig. \ref{CircuitIP}, followed by a pulsed gradient field $G_z$, which performs an
ensemble measurement of the expectation values $d_{j} = \langle\lambda_j\vert\rho_{AB}^{\varphi}\vert\lambda_j\rangle$. After the 
$(\pi/2)_{-y}$ rotations on both qubits, these $d_j$ are obtained without the need of a full state reconstruction and resulting
in the ensemble measured data $d_{j}^{exp}$.

The estimation is accomplished with an statistical estimator for $\varphi$, defined in such way that it asymptotically 
saturates the Cramer-Rao bound \cite{girolami, MGAParis}:
\begin{eqnarray}\label{estimator}
\tilde\varphi = \varphi_{0}\mathbb{I} + \frac{L_{\varphi}}{\sqrt{\nu}F(\rho_{AB}, H_{A})},
\end{eqnarray}
\par\noindent such that $\langle\tilde\varphi\rangle = \varphi_0$ and $\text{Var}(\tilde\varphi) = [\nu F(\rho_{AB}, H_{A})]^{-1}$, 
by definition. The ensemble mean and variance of this estimator are directly computed from the measured $d_{j}^{exp}$, 
the initial probe states $\rho_{AB}$ and the calculated eigenvalues $l_{\varphi}$ of $L_{\varphi}$ for each $H_A$, 
which are independent from $\varphi$.

The mean value for $\tilde\varphi$ is fitted minimizing the the function $\Theta(\varphi) = \sum_j [d_{j}^{exp} - d_{j}^{th}]^2$,
where the model $d_{j}^{th} = \langle\lambda_j \vert(e^{-i\varphi H_A}\otimes\mathbb{I}_B)\rho_{AB}
(e^{i\varphi H_A}\otimes\mathbb{I}_B)\vert\lambda_j\rangle)$ is employed. The value $\tilde\varphi$ that minimizes $\Theta$ is set as the 
expected value $\langle\tilde\varphi_{exp}\rangle$. These values are plotted in the last row of Fig. \ref{IPresults} and show good 
agreement with the true value $\varphi_0 = \pi/4$ for all settings, but the pathological one for $\rho_{AB}^{C}$ and 
$H_{A}^{(3)}$. The latter case shows how the classical probe gives an unreliable result when it commutes with the 
generating Hamiltonian.

The experimental quantum Fisher information is obtained through the expansion of $L_{\varphi}$ on its eigenbasis and using 
the measured data $d_{j}^{exp}$, as $F_{exp}(\rho_{AB}, H_A) = \sum_{j} (l_{\varphi}^{j})^2 d_{j}^{exp}$. These values are shown 
in the first row of Fig. \ref{IPresults}, along the lower bound given by $P^{A}(\rho_{AB}^{Q})$. These quantum Fisher 
information are obtained from the output of our experiments, while the interferometric power is measured on the reconstructed 
input states. For $H_{A}^{(2)}$ and $H_{A}^{(3)}$, the quantum Fisher information saturates this lower bound defined by the
interferometric power. It is also worth mentioning that $F(\rho_{AB}^{C}, H_{A}^{(3)}) = 0$, since $l_{\varphi}^{j} = 0$ 
for any $j$.

Finally, the variance of the optimal estimator is calculated replacing $\varphi_0 \mathbb{I}$ with the experimental 
mean value $\tilde\varphi_{exp}$ in Eq. (\ref{estimator}) and expanding it in terms of $l_{\varphi}^{j}$ and the measured 
$d_{j}^{exp}$ \cite{girolami}. The resulting variances are shown in the row (b) of Fig. \ref{IPresults}. This data is in 
excellent agreement with the relation $\text{Var}(\tilde\varphi_{exp} = [\nu F(\rho_{AB}, H_{A})]^{-1}$, allowing us to conclude 
that the optimal estimation strategy was performed in all settings. 

Concluding, the experimental results show that quantum discord-type quantum correlations, via the 
interferometric power, offer \emph{a priori} a minimal precision in the worst case scenario for any 
bipartite probe in a black box estimation.

\subsubsection{Final remarks}

The recognition that nuclear spins in a magnetic field would be a clear representation of qubits made NMR a 
natural candidate for quantum information processing. Indeed, because of the capability of implementing all 
basic steps necessary for QIP using conventional spectrometers, NMR was used as a bench test for most of 
the pioneering experimental demonstrations of quantum gates and algorithms. Despite that, the lack of 
scalability of pseudopure states and the criticisms concerning the absence of entanglement in room 
temperature liquid state NMR systems led to the general feeling that the contribution of the technique 
to QIP would be limited to the first demonstrations. However, this also initiated a discussion about 
the quantumness of such systems, as discussed in section \ref{NMRentang}.

Which aspects of NMR systems makes them useful in applications to quantum information experiments? 
The discovery and study of quantum correlations in separable states, the so-called general quantum 
correlations such as quantum discord, brought back the attention of the QIP community to NMR and led 
to a partial answer to the question of quantumness in NMR systems. Indeed, the facts that the system 
is highly mixed and subject to a noisy environment played in favor of NMR because these are extreme 
conditions to test the quantum properties of a system. As shown in this chapter, such correlations 
can be observed and quantified even at room temperature in a highly mixed state. As examples, we 
presented bench tests on quantifiers of general quantum correlations, discussed phenomena like the 
freezing and sudden change of quantum discord in open systems and the role of discord in guaranteeing 
a minimum precision in black box estimation in interferometry. 
 
In summary, NMR offers an excellent testbed for quantum information processing studies in few qubit 
systems, being able to perform with high precision experiments about quantum algorithms and foundations 
of quantum physics. Furthermore, the discovery of other types of quantum correlations, such as discord and 
the interest in the QIP community in understanding quantum phenomena associated to mixed states interacting 
with noisy environments brought liquid state NMR back as a method for demonstrating new concepts related 
to the quantum properties of these systems. As new perspectives, the new questions on the role of quantum 
coherences in quantum systems and in specific QIP procedures might benefit from NMR, since these concepts 
are explored since its early days.

\end{document}